\documentclass[usenatbib]{mnras}

\usepackage{amsmath}
\usepackage{graphicx}
\usepackage{mathtools}
\usepackage{cuted}
\usepackage{epstopdf}
\usepackage{ifpdf}

\def\apjl{ApJL}
\def\apjs{ApJS}
\def\aap{A\&A}
\def\mnras{MNRAS}

\title[Impact of PAH dissociation on the formation of small hydrocarbons]
{Impact of PAH photodissociation on the formation of small hydrocarbons in the Orion Bar and the Horsehead PDRs}
\author[M. S. Murga et al.]
{M. S. Murga$^{1}$\thanks{E-mail: murga@inasan.ru},
 M. S. Kirsanova$^{1,2}$, A. I. Vasyunin$^{2,3}$, Ya. N. Pavlyuchenkov$^{1}$\\
$^{1}$Institute of Astronomy, Russian Academy of Sciences, Pyatnitskaya str. 48, Moscow 119017, Russia\\
$^{2}$ Institute of Natural Sciences and Mathematics, Ural Federal University,
19 Mira Str., 620075 Ekaterinburg, Russia\\
$^{3}$ Visiting Leading Researcher, Engineering Research Institute 'Ventspils International Radio Astronomy Centre' of Ventspils University of Applied Sciences,  In\v{z}enieru 101, Ventspils LV-3601, Latvia}
\date{Accepted today. Received tomorrow; in original form \today}
\pubyear{2019}

\begin{document}
\label{firstpage}
\pagerange{\pageref{firstpage}--\pageref{lastpage}}\maketitle

\begin{abstract}
We study whether polycyclic aromatic hydrocarbons (PAHs) can be a weighty source of small hydrocarbons in photo-dissociation regions (PDRs). We modeled the evolution of 20 specific PAH molecules in terms of dehydrogenation and destruction of the carbon skeleton under the physical conditions of two well-studied PDRs, the Orion Bar and the Horsehead nebula which represent prototypical examples of PDRs irradiated by "high" and "low" ultraviolet radiation field. PAHs are described as microcanonical systems. The acetylene molecule is considered as the main carbonaceous fragment of the PAH dissociation as it follows from laboratory experiments and theory. We estimated the rates of acetylene production in gas phase chemical reactions and compared them with the rates of the acetylene production through the PAH dissociation. It is found that the latter rates  can be higher than the former rates in the Orion Bar at $A_{\rm V}<1$ and also at $A_{\rm V}>3.5$. In the Horsehead nebula, the chemical reactions provide more acetylene than the PAH dissociation. The produced acetylene participate in the reactions of the formation of small hydrocarbons (C$_2$H,  C$_3$H, C$_3$H$^{+}$, C$_3$H$_2$, C$_4$H). Acetylene production via the PAH destruction may increase the abundances of small hydrocarbons produced in gas phase chemical reactions in the Orion Bar only at $A_{\rm V}>3.5$.  In the Horsehead nebula, the contribution of PAHs to the abundances of the small hydrocarbons is negligible. We conclude that the PAHs are not a major source of small hydrocarbons in both PDRs except some locations in the Orion Bar.
\end{abstract}


\section{Introduction}
The infrared (IR) observations of the interstellar medium (ISM) reveal the wide emission bands in the wavelength range from 3 to 20~$\mu$m~\citep{gillett73, sellgren83}. It is believed that polycyclic aromatic hydrocarbons (PAHs) are responsible for arising of these bands~\citep{leger84, allamandola99}. PAHs are macromolecules consisting of benzene rings with hydrogen atoms at the periphery. The structure of these molecules may be different -- from elongated linear chains to compact PAHs. In general, the PAH molecules are planar, though, if they have defective rings consisting not of 6, but of 5 or 7 carbon atoms, the plane of the molecules can be curved.
	
While many PAHs are photostable under the conditions of the general ISM~\citep{allamandola99, tielens13}, in an environment with a high intensity ultraviolet (UV) radiation field, they can be dissociated into fragments due to the breaking of bonds as a result of interaction with high-energy photons. Photo-dissociation regions (PDRs) are examples of the objects with the increased UV radiation field. They present a transition zone between the ionized hydrogen region near a young massive star and a molecular cloud (see the detailed description in \cite{hollenbach99}). Depending on the class of the ionizing star, the intensity of the radiation field can exceed the mean value in the ISM by 1–-5 orders of magnitude. Under these conditions, PAHs dissociate and their fragments can participate in chemical reactions, in which other hydrocarbons form~\citep{allain96, verstraete01, fuente03, pety05}.

To date, a number of experiments has been carried out on the destruction of PAHs under the laboratory conditions using UV lasers ~\citep{jochims94, ekern98, zhen15}. Based on these experiments, the following conclusions can be made: 1) the main products of the dissociation are hydrogen atoms, hydrogen and acetylene (C$_2$H$_2$) molecules; 2) PAHs with a compact structure are more stable than PAHs with a non-compact structure; 3) the efficiency of the PAH destruction decreases with the increase of their size. The ionisation potential of PAHs varies depending on their  charge and size, the typical value of first ionisation potential is about 6-8~eV, while the second ionisation potential is higher, around 10-15~eV~\citep{wd01_ion}. The dissociation energies for C-H and C-C bonds typically are slightly higher than 4~eV, but they may change depending on a structure and state of PAHs. On the account of experimental data being available only for a narrow set of PAHs, theoretical modelling is necessary to estimate parameters of the destruction (rate, products, etc.) in general case. One of the theories applicable to PAHs is the “RRK” theory (Rise, Ramsperger, Kassel)~\citep{forst73}, based on which the rate of dissociation of bonds in PAHs can be estimated. Based on this theory, estimates of the time-scale of the PAH destruction in the ISM were obtained~\citep{leger89, allain96, jochims94}. In our previous studies, we have calculated the rate of the PAH destruction depending on the intensity of radiation field and PAH size~\citep{murga16a}. We have found that for PAHs with a number of carbon atoms ($N_ {\rm C}$) less than 50, when the radiation field intensity equals to or exceeds the mean value for the ISM, the time of destruction is less than 10$^5$-10$^6$~yr, which is the typical age of objects such as PDRs. PAHs with $N_ {\rm C}\approx 60-70$ can be destroyed for 10$^5$~yr but under influence of radiation field with intensity higher by 2 or more orders than the mean value for the ISM. Larger PAHs can be destroyed only partly. Thus, PAHs with $N_ {\rm C}<70$ should be destroyed in objects with an intense UV field, and this destruction should increase the number of small carbonaceous compounds available for gas phase chemistry in PDRs. 

The main channel of the dissociation of PAHs is the detachment of a hydrogen atom. An acetylene molecule detachment is less likely, however, this channel may be important for chemical composition in PDRs. Acetylene is converted through chemical reactions to different compounds such as C$_2$H and C$_3$H$_2$, which are observed in PDRs~\citep{fuente03, pety05, cuadrado15}.  To follow further evolution of acetylene molecules and their conversion to other compounds we use the chemical model MONACO which utilizes a chemical network based on OSU\footnote{A network of chemical reactions developed in the Ohio State University} and KIDA\footnote{Kinetic database for astrochemistry, {\tt\url{http://kida.astrophy.u-bordeaux.fr}}} databases~\citep{2013ApJ...769...34V, 2017ApJ...842...33V}.

To study the destruction of PAHs and its contribution to abundance of small hydrocarbons, we  consider two well-studied PDRs: the Orion Bar and the Horsehead nebula as examples of the high and low illumination PDRs. Their almost edge-on orientation \citep[][]{Walmsley2000,ODell2001} makes them suitable regions for observational and theoretical studies of the layered-on PDR structure. The Orion Bar is an ionized border of a molecular cloud irradiated by four O-B type stars from the Trapezium cluster. The most massive and luminous of them is $\theta^1$~Ori~C which spectral class is confined in the range from O5.5 class~\citep{kraus07} to O6.5V \citep{Tsivilev2014} or O7 \citep{Sota2011}. The radiation field intensity, $\chi$~\footnote{We characterize the radiation field intensity by the scale factor $\chi$, which shows how many times the intensity of the radiation field differs from the average value for the ISM described by \cite{mmp83}.}, produced by the Trapezium stars on the ionized border of the Bar at a distance of 0.235~pc is $2\cdot10^4$~\citep{goicoechea15, salgado16}. Gas density in the Bar near ionisation front is about $10^4$~cm$^{-3}$~\citep[e.g.][]{Goicoechea_2016} but then it increases and can reach $10^7-10^8$~cm$^{-3}$ in the embedded clumps \citep[][]{Leurini2010, Cuadrado2017}. Electron density ($n_{\rm e}$) in  the PDR is about $60-100$~cm$^{-3}$~\citep[see][respectively]{Mesa-Delgado2011, cuadrado19}. A parameter $\chi/n_{\rm e}$ which describes energetics and chemistry of PDRs~\citep{tielens05} achieves 10$^2$-10$^3$ by order.

The Horsehead PDR is a pillar-type dense cloud surrounded by the ionization front of IC~434 H~{\sc ii} region. The exciting star of the H~{\sc ii} region is O9.5V-type $\sigma$Ori \citep[][]{Warren1977} which gives $\chi=60$ at a projected distance of $\sim$4~pc~\citep[]{Zhou1993,Abergel2003, goicoechea09, Ochsendorf2015}. \citet{habart05} found a steep gradient of the gas density from $10^4$ to $10^5$~cm$^{-3}$ in a thin layer of 0.02~pc. Estimations of the electron density by the nebular lines are in the range from 6 to about 80-100~cm$^{-3}$ \citep[][]{Caswell1974, Bally2018}. The parameter  $\chi/n_{\rm e}$ is about 1 in the PDR region. Therefore, $\chi/n_{\rm e}$ differs at least by two orders in the considered PDRs, these regions have quite different UV-irradiation properties and energetics.

The aims of this study are: 1) to estimate the rate of the acetylene production via the PAH dissociation in two well-studied PDRs, the Orion Bar and the Horsehead nebula; 2) to compare this rate with the rate of the acetylene production in gas phase chemical reactions; 3) to estimate the contribution of acetylene produced via the PAH destruction in abundances of other small hydrocarbons. To perform this, we consider in detail the evolution of PAHs of different sizes under the physical conditions of PDRs in terms of their destruction and acetylene formation, and we apply a chemical model of the acetylene formation to the same conditions. We also estimate the abundance of small hydrocarbons with the contribution of PAHs and without it in order to understand, whether PAHs can be their weighty source in PDRs. 

\section{The model of PAH evolution} \label{model}

In this work we consider a history of state of specific PAH molecules under PDR conditions. The state is characterized by the number of carbon and hydrogen atoms in a molecule. The processes of dissociation (hydrogen and carbon atoms loss) and hydrogenation are considered, but the processes of formation and growth of PAHs are out of the scope of this work. As mentioned above, the experiments show that a hydrogen atom is the most frequent product of the PAH dissociation, while an acetylene molecule is produced with less efficiency~\citep{allain96, ekern98, berne12,  zhen14a, zhen15, joblin19}. It was shown in  \cite{allain96} that the rate of H-loss due to PAH photodissociation exceeds the rate of C$_2$H$_2$-loss by 1-2 orders. In accordance to this, generally, PAHs became partly or even fully dehydrogenated first, and after that they lose carbon atoms. Obviously, that fully dehydrogenated PAHs can not produce acetylene any more, only diatomic carbon. However, hydrogen atoms can be attached to PAHs again, and the number of hydrogen atoms in molecules can be partially  restored. Thus, in order to calculate the number of produced acetylene molecules by PAHs correctly, it is necessary to take into account the hydrogenation state of molecules.
 
Let an $i$th molecule with charge $Z$ loses hydrogen atoms, molecules of hydrogen and acetylene (or diatomic carbon) with the rates $R_{\rm H}^{i, Z}$, $R_{\rm{H}_2}^{i,Z}$, $R_{\rm{C}_2\rm{H}_2}^{i,Z}$, atoms~s$^{-1}$, correspondingly (the rates for molecules are multiplied by 2 to get the units of atoms~s$^{-1}$), while hydrogen atoms and molecules attach to the molecule with the rate $H^{i,Z}$, atoms~s$^{-1}$. The methods for calculating these rates are described below. Changes of the number of carbon and hydrogen atoms for the $i$th molecule  ($N_{\rm C}^{i,Z}/dt$ and $N_{\rm H}^{i,Z}/dt$, correspondingly) are calculated from the equations
\begin{eqnarray} \label{dif_eq}
\frac{dN_{\rm C}^{i,Z}}{dt} &=& - 2R_{{\rm C}_2{\rm H}_2}^{i,Z}\\ \nonumber
\frac{dN_{\rm H}^{i,Z}}{dt} &=& - R_{\rm H}^{i,Z} - 2R_{\rm{H}_2}^{i,Z}-2R_{{\rm C}_2{\rm H}_2}^{i,Z}+H^{i,Z}. 
\end{eqnarray}

We use the explicit Euler method to solve the system of equations~\ref{dif_eq}, therefore the number of carbon and hydrogen atoms at the $j$th   time step, $t_{j} = t_{j-1} + \Delta t$, can be expressed through
\begin{eqnarray}\label{int_eq}
N_{\rm C}^{i,Z}(t_j) &=& N_{\rm C}^{i,Z}(t_{j-1}) + \frac{dN_{\rm C}^{i,Z}}{dt} \Delta t\\ \nonumber
N_{\rm H}^{i,Z}(t_j) &=& N_{\rm H}^{i,Z}(t_{j-1}) + \frac{dN_{\rm H}^{i,Z}}{dt} \Delta t, 
\end{eqnarray}
where $\Delta t$ is an integration time step. The index $j$ starts at 1, and $t_0=0$~yr. The value of $\Delta t$ is chosen equal to 1~yr, as decreasing it lower than this value does not change the final result. The number of acetylene molecules produced by the $i$th PAH molecule during $N_{\rm iter}$ time steps is estimated as
\begin{eqnarray}
N_{\rm{C}_2\rm{H}_2}^{i,Z} = \frac{1}{2}\sum\limits_{j=1}^{N_{\rm iter}} \left(N_{\rm C}^{i,Z}(t_j)-N_{\rm C}^{i,Z}(t_{j-1})\right), \mbox{if}\; N_{\rm H}^{i,Z}(t_{j-1})>1.
\label{n_acet}
\end{eqnarray}

The condition $N_{\rm H}^{i,Z}(t_{j-1})>1$ indicates that the PAH can not produce acetylene during the integration step, in which it is completely dehydrogenated or has only one hydrogen atom. However, owing to reverse process, hydrogenation, acetylene can be obtained from the molecule again in subsequent steps. The number of integration steps for a particular molecule depends on what comes first: the molecule completely loses all carbon atoms or the maximum computation time is reached, which is assumed to be $10^6$~yr as it is the characteristic age of massive stars, around which PDRs appear.

In the following text, we will use the terms “normal", “dehydrogenated” and “super-hydrogenated" states. A PAH in the normal state has the same ratio between carbon and hydrogen atoms ($X_{\rm H}$) as in its original state ($X_{\rm H}^{0}$), i.e. each peripheral carbon atom is connected to one hydrogen atom. If this ratio becomes greater than the initial one, then the state is considered as super-hydrogenated or saturated by hydrogen, if less, then the state is dehydrogenated.

\subsection{Dissociation rate of PAHs}\label{dissoc}

In this work, the rates of PAH dissociation ($R_{\rm H}^{i,Z}$, $R_{\rm{H}_2}^{i,Z}$, $R_{\rm{C}_2\rm{H}_2}^{i,Z}$) are calculated according to the scheme presented in the works of \cite{murga16a, murga19} with some modifications described below. The photon energy obtained by a PAH molecule can lead to several processes: 1) emission of an IR photon; 2) PAH ionization; 3) bond dissociation. These processes compete each other, and the probability of each of them depends on the photon energy, size and ionization state of PAH. The smaller a PAH is, the more likely its dissociation becomes, with increasing PAH size the probabilities of the ionization and emission of an IR photon increase as well. 

We assume that the photon energy ($E$) either is spent to ionization or converts to the internal energy of a PAH ($E_{\rm in}$). We adopt the ionization yield (or probability of ionization) from \cite{wd01_ion}. The probability of dissociation of a PAH of a radius $a_i$ and charge number $Z$ with the internal energy $E_{\rm in}$ can be found as
\begin{equation}
Y_{\rm diss}^{\rm sp}(a_i,E_{\rm in}, Z) = \frac{k_{\rm diss}(E_{\rm in})}{k_{\rm diss}(E_{\rm in})+k_{\rm IR}(a_i,E_{\rm in},Z)},
\end{equation}
where  $k_{\rm diss}$ and $k_{\rm IR}$ are the bond dissociation and the IR emission rates correspondingly.

In our previous paper~\citep{murga16a}, we have concluded that PAHs with $N_{\rm C}>70-80$ are photostable, the time of their dissociation is much longer than the lifetime of the objects where they are observed. In this work, the updated model of the PAH destruction is used, namely the multi-photon heating mechanism of PAHs has been added~\citep{guhathakurta89}. Thus, we take into account that a PAH can accumulate the energy obtained from several photons, if it does not cool down due to IR-photon emission in the time interval between the entries of these photons. Therefore, the PAH internal energy can be higher, and, as a result, the rate of the destruction of PAHs increases compared to the rate calculated by considering only single-photon heating. The multi-photon mechanism is mostly important for large PAHs ($N_{\rm C}>80$), which can also be destroyed in PDRs due to the high intensities of radiation field. For each PAH, we find the probability distribution of internal energies, $p(a_i, E_{\rm in})$, that it may have. To estimate the total probability of the dissociation of this PAH, we integrate probabilities of the dissociation corresponding to the specific internal energies, taking into account the probability of being in the states with these internal energies, i.e.
\begin{equation}
Y_{\rm diss}(a_i, Z) = \int\limits_{0}^{\infty}Y_{\rm diss}^{\rm sp}(a_i,E_{\rm in}, Z)p(a_i, E_{\rm in})dE_{\rm in}.
\end{equation}

The final rate of dissociation is found taking into account the PAH absorption cross section $C_{\rm abs}(a_i,E, Z)$ and the photon flux of radiation field, $F(E)$, as
\begin{equation}
R_{\rm frag}^{i,Z} = Y_{\rm diss}(a_i, Z)\int\limits_{5~\rm{eV}}^{13.6~\rm{eV}}  C_{\rm abs}(a_i,E, Z) F(E) dE.
\end{equation} 
The photon energies from 5 to 13.6~eV are considered as relevant for the destruction and ionisation processes. Depending on which fragment of dissociation is considered (H, H$_2$ or C$_2$H$_2$), the rate $R_{\rm frag}^{i,Z}$ is designated as $R_{\rm H}^{i,Z}$, $R_{\rm{H}_2}^{i,Z}$, $R_{\rm{C}_2\rm{H}_2}^{i,Z}$, correspondingly. $C_{\rm abs}(a_i,E, Z)$ for PAHs is taken from the work of \cite{DL07}. The photon flux of radiation field in units eV$^{-1}$~cm$^{-2}$~s$^{-1}$ is obtained from the radiation field intensity taken from \cite{mmp83} scaled by factor $\chi$ which precise value for a specific PDR is described below.

In contrast to the previous works \citep{murga16a, murga19}, the PAH is described using the Gibbs microcanonical distribution, and its bond dissociation rate is estimated using the expression in the form of the Arrhenius law (for more details, see \cite{tielens05}:
\begin{equation}
k_{\rm diss}(E_{\rm in}) = k_0(T_e)\exp\left(-\frac{E_0}{k_{\rm B}T_e}\right), 
\end{equation}
where $k_{\rm B}$ is the Boltzmann constant, $E_0$ is the activation energy of a bond, $T_e$ is the characteristic effective temperature of PAH associated with the microcanonical temperature $T_{\rm m}$ and the internal energy through
\begin{equation}
T_e = T_{\rm m}\left(1-0.2\frac{E_0}{E_{\rm in}}\right),
\end{equation}
and $k_0$ is a coefficient estimated by
\begin{equation}
k_0(T_e) = \frac{k_{\rm B}T_e}{h}\exp\left(1+\frac{\Delta S}{\mathcal R} \right),
\end{equation}
where $h$ is the Planck constant, $\Delta S$ is the change of entropy due to the bond dissociation, and ${\mathcal R}$ is the universal gas constant. We use the relationship between the microcanonical temperature and the internal energy of PAH presented in the work of \cite{DL01_stoch}. The parameters $E_0$ and $\Delta S$ are determined by the type of the bond and the state of the PAH (charge and hydrogenation state). The values adopted in this work are presented in Table~\ref{tab: E0,S} and shortly described below. 

Regarding to the normal and dehydrogenated states, PAHs with even and odd numbers of hydrogen atoms have different $E_0$ and $\Delta S$ in the same hydrogenated state according to \cite{andrews16}. But in this work, we perform calculations with no consideration of parity. So we take the mean values between the ones for PAHs with even and odd number of hydrogen atoms in the normal and dehydrogenated states given by \cite{andrews16}, namely, $E_0 = 4.3$~eV and $\Delta S = 11.8$~kcal~K$^{-1}$~mol$^{-1}$, regardless of the charge and parity of the number of hydrogen atoms. According to the same paper, we adopt the values $E_0 = 3.52$~eV and $\Delta S = -12.69$~kcal~K$^{-1}$~mol$^{-1}$ for the loss of the hydrogen molecule. For the acetylene molecule, the values are determined less precisely. Based on the works of \cite{mic10} and \cite{ling98}, we take values 4.6~eV and 10~kcal~K$^{-1}$~mol$^{-1}$ for $E_0$ and $\Delta S$, correspondingly.

If the PAH is super-hydrogenated, then it may be less stable than its less hydrogenated analogue, as indicated by the experiments~\citep{wolf16, quitian18, rapacioli18}. Fragmentation of such molecules via detachment of carbonaceous compounds becomes the dominant channel. There can be a variety of C$_{x}$H$_{y}$ among the compounds, not just the acetylene molecule. Nevertheless, in this paper, we consider the acetylene molecule as the only carbonaceous product of PAH destruction. In the absence of estimates of $\Delta S$ for acetylene in the super-hydrogenated state, we adopt the same value as for the normal and dehydrogenated states. From the experiment of \cite{wolf16} it can be determined that the activation energy for acetylene becomes less than 2.73~eV, since only one photon with such energy is required for dissociation, but the precise value has not been measured. The activation energy for the dominant carbonaceous channel (CH$_3$) is estimated to be about 2~eV (the exact value depends on the fraction of extra hydrogen atoms in the PAH). So we accept $E_0 = 2$~eV for acetylene as we consider only it among carbonaceous products. For the loss of the hydrogen atoms in the super-hydrogenated state we use the values adopted  in the work of \cite{andrews16}, namely, $\Delta S = 13.3$~kcal~K$^{-1}$~mol$^{-1}$, $E_0 = 1.4$~eV ($Z\leq 0$) and $E_0 = 1.55$~eV ($Z>0$). In the super-hydrogenated states, the hydrogen molecule can be detached from the PAH due to the Eley-Rideal mechanism; therefore, we do not consider this channel as a competitor in the dissociation. Instead, similarly to the work of \cite{andrews16}, we calculate the additional hydrogen loss with the rate of $k_{\rm er}$ equal to
\begin{equation}
k_{er} = 8.7\cdot10^{-13}\sqrt{\frac{T_{\rm gas}}{100}}n_{\rm H}, 
\end{equation}
where  $T_{\rm gas}$ -- the gas temperature and $n_{\rm H}$ -- the atomic hydrogen number density. 

\begin{table}
\caption{Dissociation parameters $E_0$~[eV] and $\Delta S$~[cal~K$^{-1}$~mol$^{-1}$].}
\label{tab: E0,S}
\begin{center}
\begin{tabular}{|l|c|c|c|c|}
\hline
Fragment & \multicolumn{2}{c|}{$X_{\rm H}\leq X_{\rm H}^{0}$} & \multicolumn{2}{c|}{$X_{\rm H}> X_{\rm H}^{0}$} \\
\hline
          &   $E_0$       &     $\Delta S$     &   $E_0$      &       $\Delta S$       \\
\hline
H  ($Z\leq0$)        &     4.3                &       11.8              &     1.4             &   13.3    \\
 H  ($Z>0$)     &        4.3            &        11.8            &     1.55               &   13.3   \\
H$_2$&      3.52             &      -12.69           &        -            &    -   \\
C$_2$H$_2$&     4.6      &       10.0              &      2.0             &    10.0    \\
\hline
\end{tabular}
\end{center}
\end{table}

\subsection{Hydrogenation of the PAH molecules}\label{acrretion}
The total rate of the hydrogenation  of PAHs (or H-addition) can be expressed as follows
\begin{equation}
H^{i,Z} =
\left\{
\begin{array}{ll}
n_{\rm H} k^{i,Z=0},\mbox{if}\; Z=0 \\
n_{\rm H}k^{i,Z\neq0}+n_{{\rm H}_2}k^{i,Z\neq0}_{{\rm H}_2},\mbox{if}\; Z\neq0
 \end{array} \right .
\end{equation}
where $k$ are the rates of H-addition for the different cases which are described below.  

\begin{figure*}
	\includegraphics[width=0.8\textwidth]{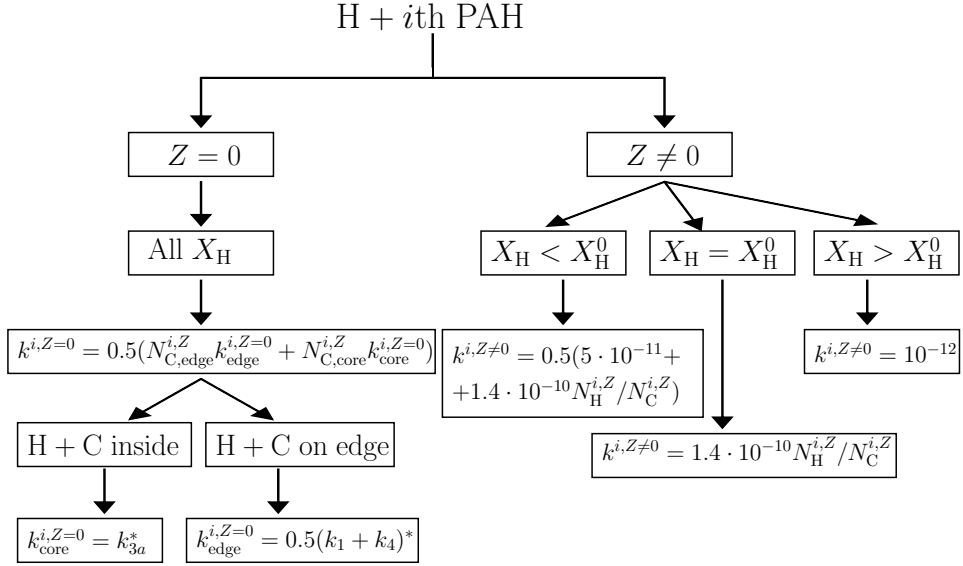}
	\caption{Scheme with the rates of the H-addition reaction for different PAH states. $k^{i,Z=0}$ and $k^{i,Z\neq0}$ are given in units cm$^3$~s$^{-1}$. $^*$We designate different points of attachment of hydrogen atom to PAH analogically to the paper of \protect\cite{goumans11}. Specifically, they performed calculations for the points with designations 1, 3a and 4.}
	\label{block}
\end{figure*}

The rates of the hydrogenation differ for PAHs of different sizes, hydrogenation states and charges. We schematically illustrate all the states and the corresponding expressions for the rates in Fig.~\ref{block}. We adopt the rates for ionized PAHs from the work of \cite{montillaud13}. Specifically, if PAHs are in the normal state, then the rate of the H-addition, $k^{i, Z\neq0}$, equals to 1.4$\cdot10^{-10}$~cm$^3$~s$^{-1}$ scaled by the ratio $N_{\rm H}^{i,Z}$/$N_{\rm C}^{i,Z}$. For PAHs in the super-hydrogenated state the reaction rate is $10^{-12}$~cm$^3$~s$^{-1}$ regardless of $N_{\rm H}^{i,Z}$ and $N_{\rm C}^{i,Z}$. Finally, if PAHs are dehydrogenated, the reaction rate is taken as the mean between the values for PAHs with the odd and even number of carbon atoms given in the work of \cite{montillaud13}: $k^{i, Z \neq0} = 0.5 (1.4 \cdot10^{-10} N_{\rm H}^{i,Z}/N_{\rm C}^{i,Z} +5 \cdot10^{-11})$~cm$^3$~s$^{-1}$. The reaction rate with the molecular hydrogen, $k^{i, Z \neq0}_{\rm{H}_2}$, is adopted to be equal to 5$\cdot10^{-13}$~cm$^3$~s$^{-1}$ for all ionized PAHs in any hydrogenation state.

In the papers of \cite{lepage01, lepage03, montillaud13} the reactions of the H-addition to neutral PAHs were neglected. However, it was shown that neutral PAHs can also efficiently react with hydrogen atoms~\citep{rauls08}. In the work of \cite{goumans11}, the calculations of the chemisorption process of hydrogen atoms on PAHs were carried out by considering the quantum mechanical tunnelling through reaction barriers; the rates of the reaction were calculated depending on temperature and the point of attachment (core or periphery). In the same work of \cite{goumans11} it was shown that this process should not be neglected, since it is probably a key process to the formation of the H$_2$ molecule and super-hydrogenated PAHs. 

\cite{goumans11} considered three locations of the attachment of H atoms to PAHs, two peripheral and one central, which they designate by numbers 1, 4 and 3a, correspondingly. The reaction rates for the peripheral carbon atoms are much higher than those for the inner atoms, so it is important where the hydrogen atom is attached. As the rates are given per site, we need to consider how many carbon atoms are on the edge and inside a PAH. For the initial state, the number of the peripheral carbon atoms, $N_{\rm C, edge}^{i, Z}(t_0)$, is taken equal to the number of the hydrogen atoms in a PAH, $N_{\rm H}^{i, Z}(t_0)$, the rest are considered as the central ones. That is, the number of the central atoms can be found as $N_{\rm C}^{i, Z}(t_0) -N_{\rm H}^{i, Z}(t_0)$. For simplicity, the ratio between the number of the central and the peripheral atoms is assumed to be constant, regardless of how much the PAH has lost during its evolution. Thus, at each time step, the number of carbon atoms at the edge and in the center ($N_{\rm C, edge}^{i,Z}(t_j)$ and $N_{\rm C, core}^{i,Z}(t_j)$, correspondingly) are calculated based on the initial ratio, and the current number of carbon atoms. The rate of the H-addition to the peripheral carbon atoms of neutral PAHs, $k^{i, Z = 0}_{\rm edge}$, is considered as the mean value between the rates of two peripheral points considered in the work of \cite{goumans11} (with indices 1 and 4). For the central carbon atoms, we adopt the reaction rate, $k^{i, Z = 0}_{\rm core}$, which corresponds to the point 3a from the work of \cite{goumans11}. The total rate for neutral PAHs is estimated as the mean value between the rates for peripheral and inner atoms. It should be noted that the reaction rate significantly depends on the gas temperature, and when the temperature is more than 1000~K, the value of the rate considerably exceeds the values for ionized PAHs. Such temperatures can be achieved in PDRs close to the region of ionized hydrogen. Thus, it turns out, neutral PAHs under these conditions can be super-hydrogenated.

Due to the lack of data about the reaction between neutral PAHs and the molecular hydrogen, we do not consider this reaction. Moreover, the hydrogenation of ionized PAHs by molecular hydrogen is, in fact, ineffective, and it does not affect the results of calculating the rate of the PAH dissociation and the formation of acetylene~\citep{montillaud13}.

\section{Acetylene production via PAH dissociation}\label{results}

We have selected 20 compact PAH molecules to calculate the yield of acetylene. These PAHs chosen are the most abundant in the ISM according to the recent work of \cite{andrews15}: C$_{16}$H$_{10}$, C$_{30}$H$_{14}$, C$_{42}$H$_{24}$, C$_{48}$H$_{18}$, C$_{96}$H$_{26}$ (pyrene family), C$_{24}$H$_{12}$, C$_{54}$H$_{18}$, C$_{96}$H$_{24}$, C$_{150}$H$_{30}$ (coronene family), C$_{32}$H$_{14}$, C$_{66}$H$_{20}$, C$_{112}$H$_{26}$ (ovalene family), C$_{40}$H$_{16}$, C$_{78}$H$_{22}$, C$_{128}$H$_{28}$  (anthracene family), C$_{48}$H$_{12}$, C$_{90}$H$_{24}$, C$_{144}$H$_{30}$ (tetracene family), C$_{110}$H$_{26}$, C$_{130}$H$_{28}$ (some large compact PAHs). Differences in size, number of benzene rings and hydrogen atoms, structure and ratio between central and peripheral carbon atoms lead to different rates of acetylene production by these molecules under the same conditions.

\begin{figure*}
	\includegraphics[width=0.45\textwidth]{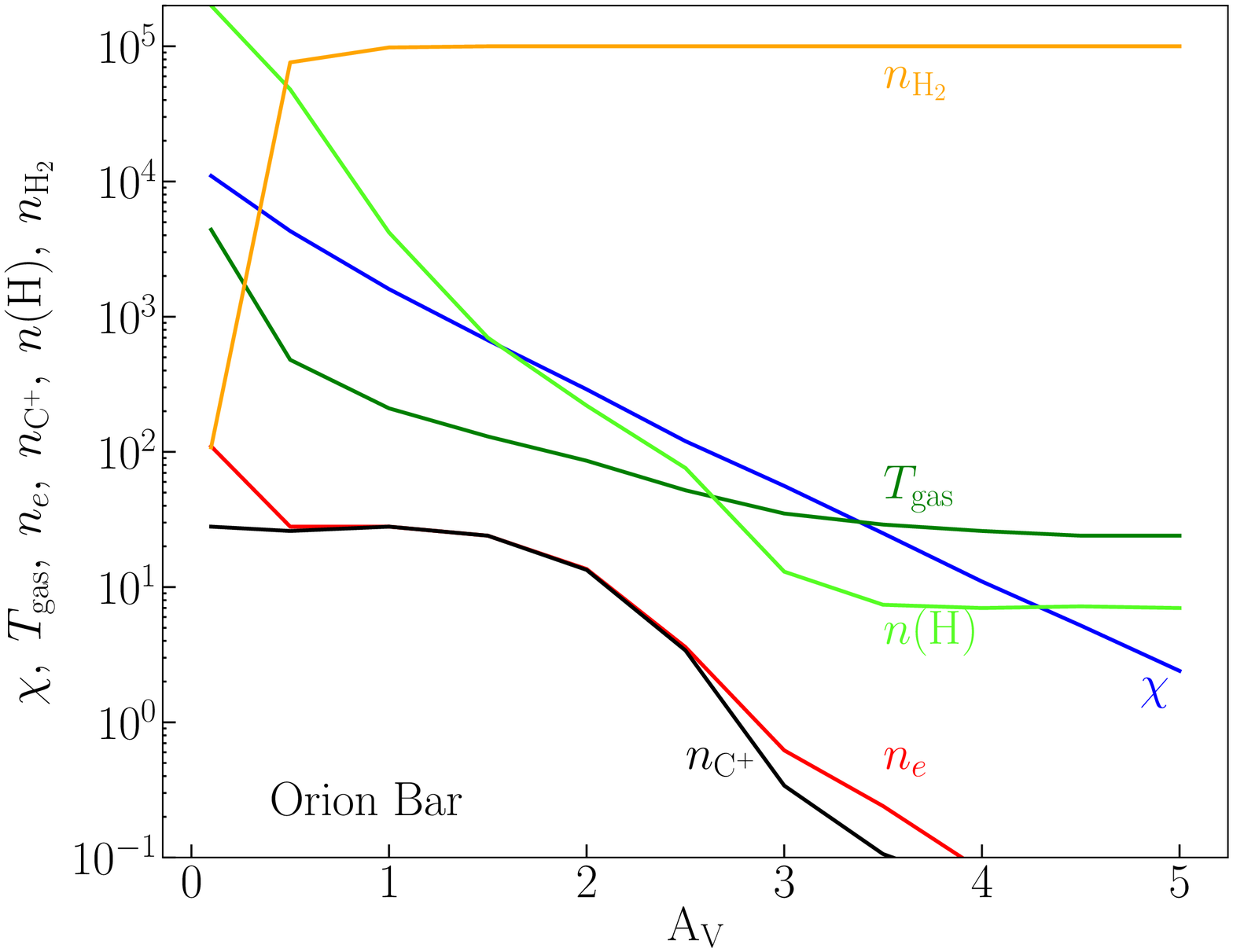}
	\includegraphics[width=0.45\textwidth]{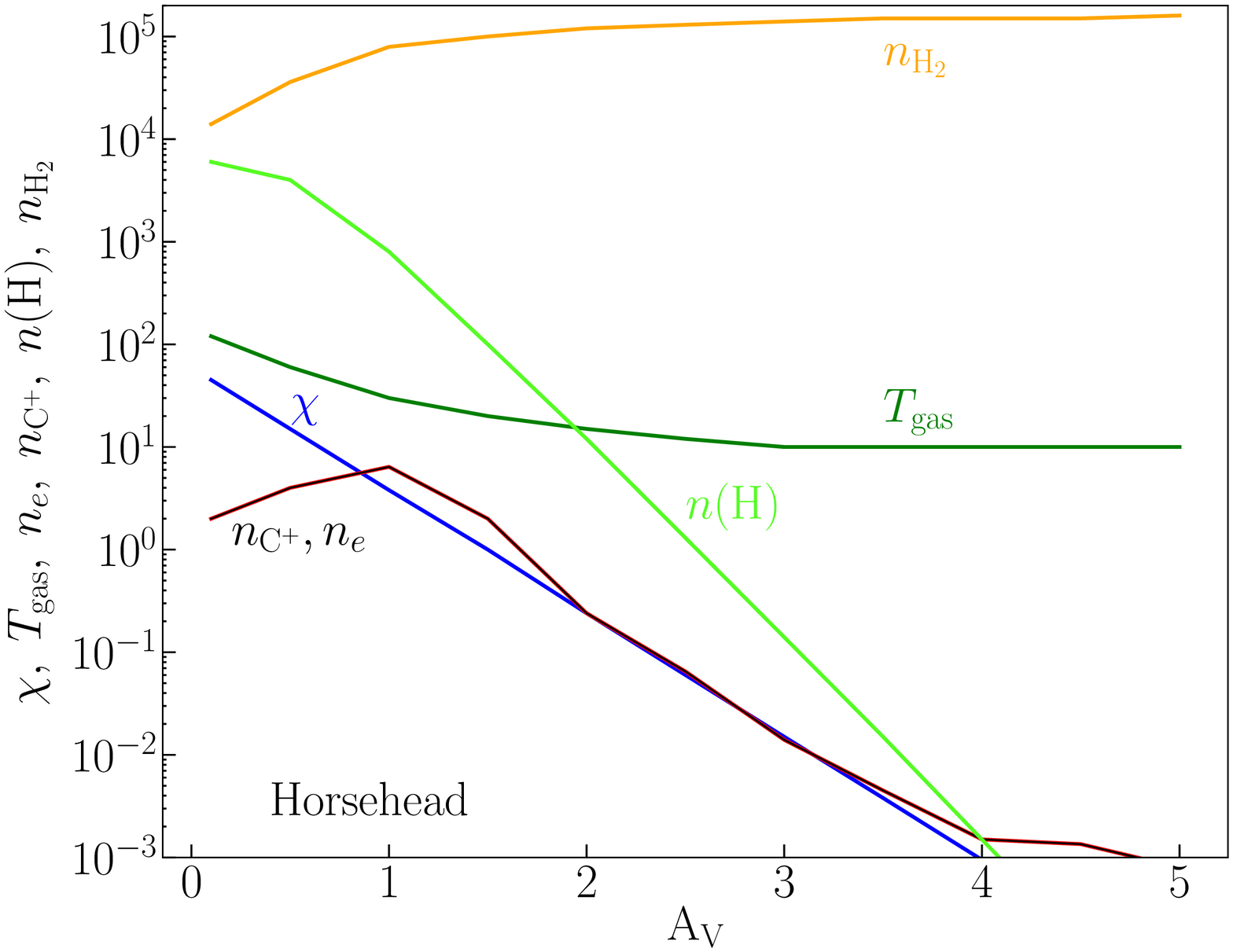}
	\caption{Profiles of the physical parameters in PDRs of the Orion Bar in approach of the isochoric model~\citep{goicoechea15}~(on the left) and the Horsehead nebula~\protect\citep{goicoechea09, legal17}~(on the right) which were adopted for calculations in this work. Temperature is given in K, densities are given in cm$^{-3}$.}
	\label{profiles}
\end{figure*}

We investigate two well-studied PDRs, the Orion Bar and the Horsehead nebula which were shortly described above. We take the physical conditions at different distances from the ionizitaion front. For our calculations, it is necessary to have the following set of parameters: the number densities of atomic hydrogen ($n_{\rm H}$) and molecular hydrogen ($n_{{\rm H}_2}$), electrons ($n_{\rm e}$) and carbon ions ($n_{{\rm C}^{+}}$), the gas temperature, and the radiation field intensity. For the Orion Bar the profiles of parameters along a distance from the ionization front are taken from \cite{goicoechea15} where the isochoric (constant-density) model was considered, and from \cite{goicoechea19} where the isobaric (constant-pressure) model was considered. In the main text we present results for the first model, and the results for the second one are given in \ref{app}. For the Horsehead nebula the profiles of $n_{\rm H}$, $n_{\rm e}$, $n_{{\rm C}^{+}}$ are taken from the work of \cite{goicoechea09}, the profiles of $n_{{\rm H}_2}$ and $T_{\rm gas}$ are from the work of \cite{legal17}. The intensity of the radiation field relative to the value near the ionizing source ($\chi_0$) is determined through the absorption in the V band as
\begin{equation}
c^{\rm UV}_{\rm V}A_{\rm V} = -2.5\log_{10}(\chi/\chi_0)
\end{equation}
where $A_{\rm V}$ is the extinction value and $c^{\rm UV}_{\rm V}$ is a coefficient between extinction in the UV range and V band. Using the correlations suggested by \cite{cardelli89}, we calculated the mean value in the wavelength range which is considered in the PAH dissociation model (912 -- 2500~\AA{}). It is approximately equal to 3. We use the value $\chi_0 = 60$, which is often adopted for the calculations in this PDR~\citep{goicoechea09, legal17, rimmer12}. In Fig.~\ref{profiles} the profiles of the physical conditions for two PDRs are presented. We note that it is supposed that the ionisation fronts are located at $A_{\rm V}=0$, and the minimum distance in our calculations corresponds to $A_{\rm V}=0.1$.

The rates of the PAH destruction and the hydrogenation depend on charge of the PAH as it was mentioned above. The charge was calculated based on the work of \cite{wd01_ion}, where it is determined by interaction with the UV radiation field, ions (in our case, C$^{+}$) and electrons. In fact, the charge state of the $i$th PAH is described by the probability function, $f(i,Z)$, value of which we estimate for each charge number Z. Further, we take this function into account for calculating the number and the production rate of acetylene molecules under certain conditions.
 
\begin{figure*}
	\includegraphics[width=0.45\textwidth]{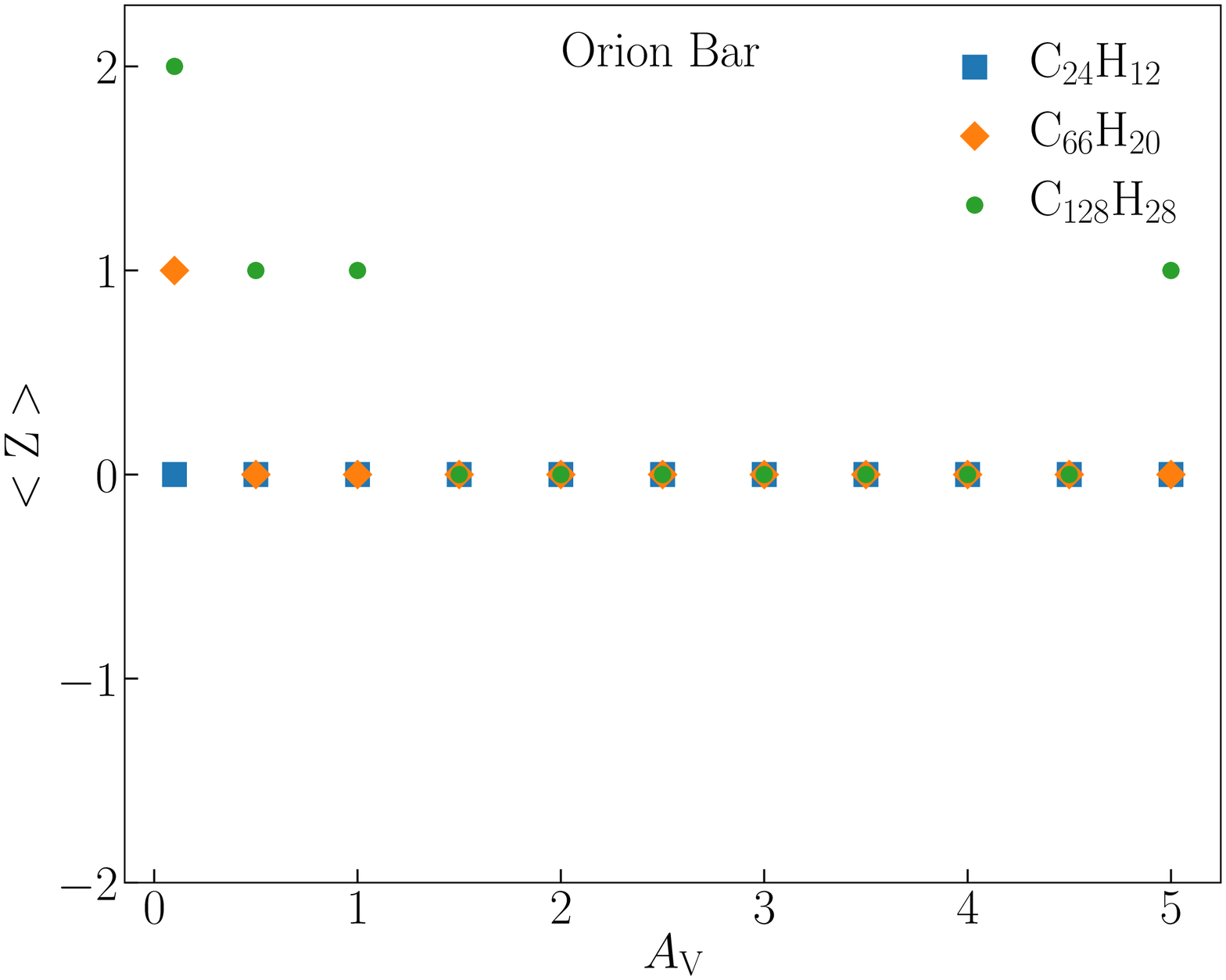}
	\includegraphics[width=0.45\textwidth]{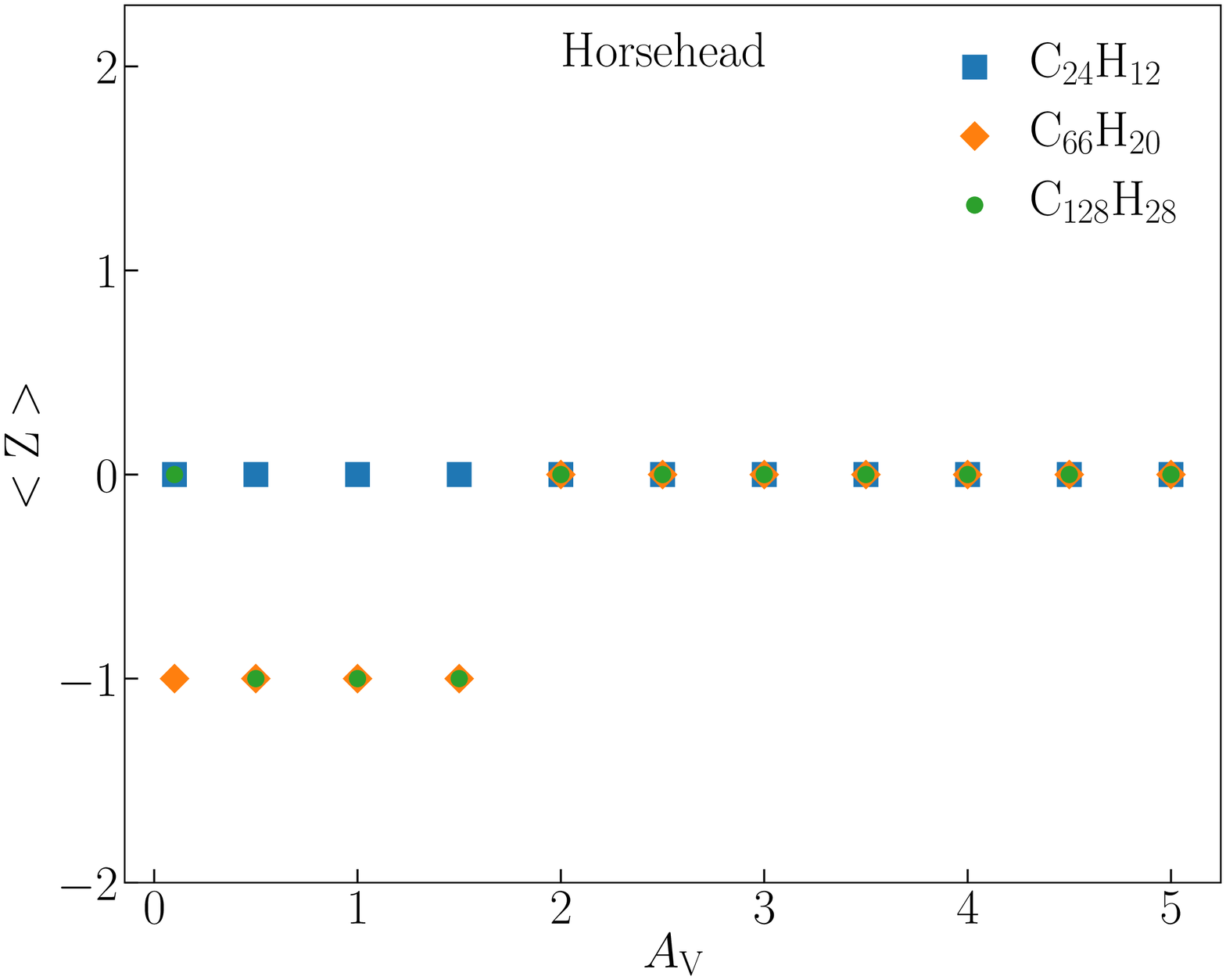}
	\caption{Dependence of the mean charge numbers on position in PDRs (on the left -- the Orion Bar, on the right -- the Horsehead nebula) for three PAHs considered in the work.}
	\label{charges}
\end{figure*}

In Fig.~\ref{charges} the mean charge numbers for three PAHs (C$_{24}$H$_{12}$, C$_{66}$H$_{20}$, C$_{128}$H$_{28}$) along PDRs are illustrated. We have chosen these PAHs as representative of small, medium and large PAHs. For the Orion Bar, where the UV field is by 2 orders higher than in the Horsehead nebula, PAHs can be both neutral and positively charged, large PAHs are ionized twice. In the Horsehead nebula, small PAHs are neutral, while the larger ones can have also a negative charge number in the extinction range $0.1<A_{\rm V}<2$. 

\begin{figure*}
	\includegraphics[width=0.45\textwidth]{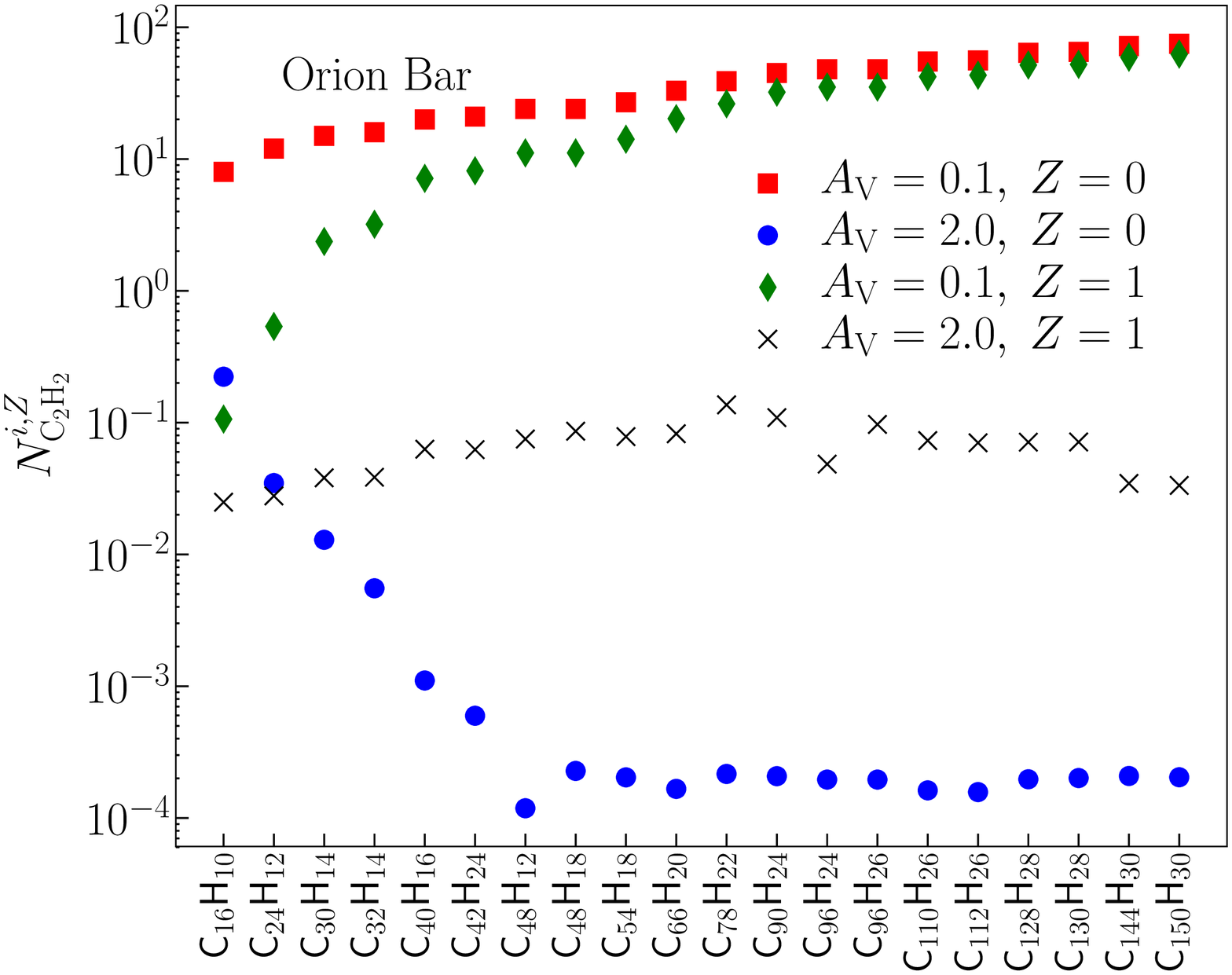}
	\includegraphics[width=0.45\textwidth]{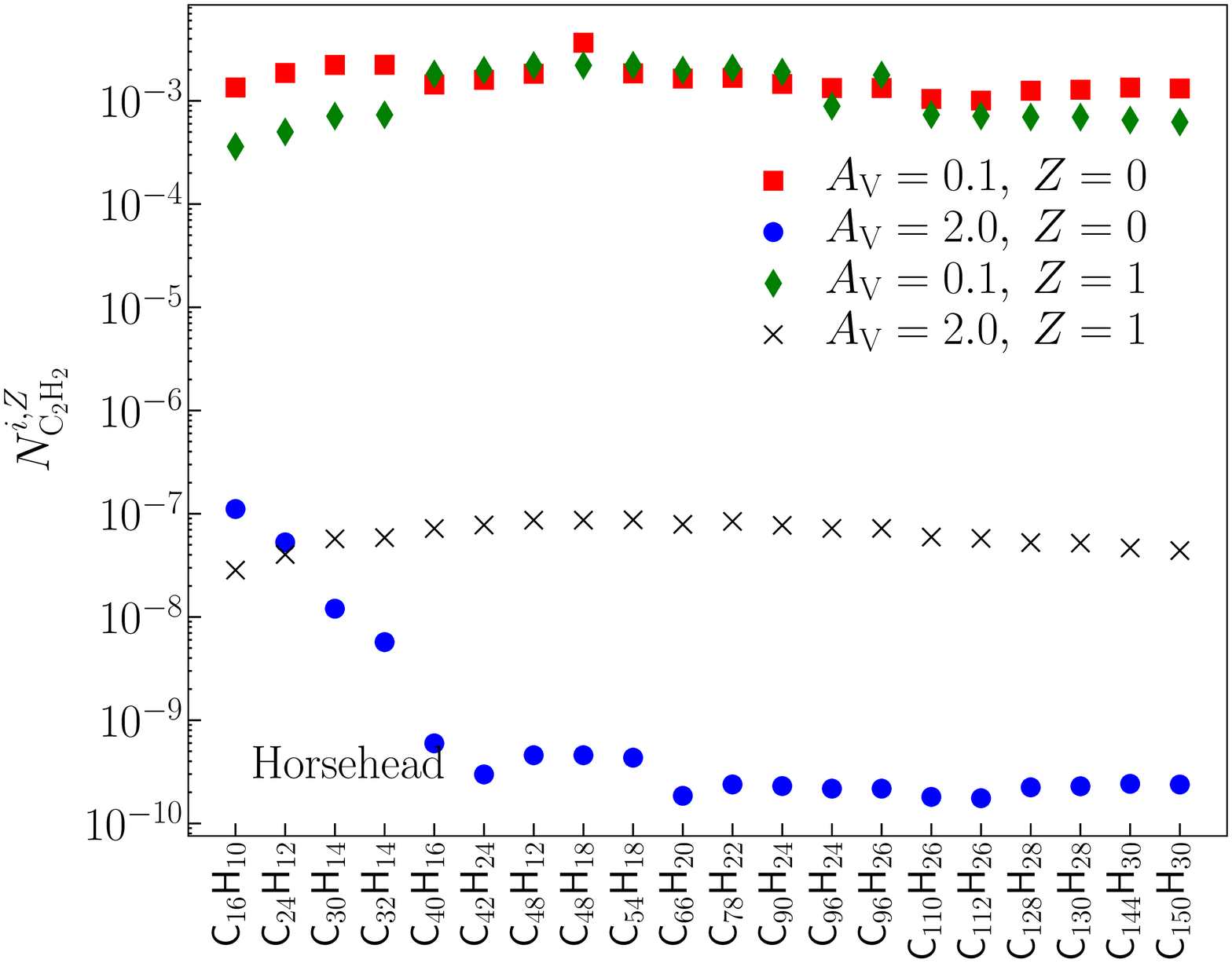}
	\caption{The number of the produced acetylene molecules by each PAH molecule for the different physical conditions in the Orion Bar in approach of the isochoric model (on the left) and in the Horsehead nebula (on the right).}
	\label{nc2h2_mol}
\end{figure*}

In Fig.~\ref{nc2h2_mol} the numbers of the acetylene molecules produced by specific PAHs under the conditions of $A_{\rm V}=0.1$ and $A_{\rm V}=2$ for both considered PDRs are shown. It is seen that the ionized and neutral PAHs produce different numbers of acetylene molecules, and the amount of the produced acetylene varies depending on conditions at the same ionization state. The rate of the hydrogenation of neutral PAHs sharply increases with temperature, when the temperature exceeds $\sim$200~K, and it also depends linearly on the hydrogen density. The adopted rates of hydrogenation  of ionized PAHs do not correlate with temperature, only with density (linearly, as well). In the Orion Bar at $A_{\rm V}=0.1$ the gas temperature is more than 1000~K. According to \cite{goumans11}, under these conditions the hydrogenation process of neutral PAHs is very efficient and may compensate the H-loss, therefore, neutral PAHs of all sizes are almost always in the normal or in the super-hydrogenated states, i.e. they can produce acetylene. Ionized PAHs with $N_{\rm C}> 70$ produce a comparable number of acetylene under these conditions, smaller PAHs produce less by 1-2 orders, because the H-loss process for smaller PAHs dominates over hydrogenation. Thus, small ionized PAHs produce acetylene slower in comparison with the larger ones. 

At $A_{\rm V}=2$ the gas temperature and the density of atomic hydrogen decrease, and due to the strong dependence of the hydrogenation rate of neutral PAHs on temperature, it becomes much less under these conditions than in the previous case. The radiation field remains sufficient for efficient dehydrogenation. So, the neutral PAHs quickly lose their hydrogen atoms, and they do not have time to produce enough acetylene. Roughly speaking, the number of the produced acetylene molecules by neutral PAHs is proportional to their dissociation rate, hydrogenation can be neglected. Small neutral PAHs with $N_{\rm C}<40$ produce higher amount of acetylene than larger ones. That is because their rates of H-loss and C$_2$H$_2$-loss do not differ from each other so much as the rates for larger PAHs. Large neutral PAHs become dehydrogenated when they have produced just a small amount of  acetylene (rate of the H-loss is much higher than the rate of C$_2$H$_2$-loss). Due to the low rate of hydrogenation they can no longer produce acetylene after losing their H atoms. The acetylene production rate for ionized PAHs is not straightforward proportional to the dissociation rate, because their hydrogenation rate is not sensitive to temperature and decreases less steeply than the rate for neutral PAHs.  

The discrepancy in the numbers of produced acetylene molecules under the different radiation field intensities is also observed in the Horsehead nebula (in Fig.~\ref{nc2h2_mol}). The field in this PDR is not as intense as in the Orion Bar, so the number of the produced acetylene molecules is less than in the Orion Bar. Nevertheless, it can also be seen that at $A_{\rm V}=0.1$ (a higher radiation field intensity, a higher temperature and atomic hydrogen density) PAHs produce much more acetylene than at lower values of these parameters. This fact reflects that $n_{\rm H}$ becomes negligible for large values of $A_{\rm V}$ (and low radiation field intensities), and, as a result, hydrogenation does not occur. Like the Orion Bar, there is a difference between the numbers of acetylene produced by PAHs with different charges under the same conditions. The manner of the difference is approximately the same in spite of the discrepancy between the radiation field intensities. However, it is seen that the amount of produced acetylene molecules is much lower than in the Orion Bar.

Summarizing the results presented in Fig.~\ref{nc2h2_mol}, it can be concluded that the efficiency of the acetylene production is sensitive to many parameters simultaneously: the size and the charge of PAH, the intensity of the radiation field, the atomic hydrogen density and gas temperature. It is hardly possible to highlight which PAHs or conditions are the best for acetylene production, because all the parameters should be considered together.

In order to follow the production of acetylene at various distances from the ionization front in PDRs, we integrate the values of $N_{{\rm C}_2 {\rm H}_2}^{i, Z}$ over all the ionization states using the probability function $f(i,Z)$ and we get the value $N_{{\rm C}_2{\rm H}_2}^{i}$ for each $i$th molecule, i.e.
\begin{equation}
N_{{\rm C}_2{\rm H}_2}^{i} = \int N_{{\rm C}_2 {\rm H}_2}^{i, Z}f(i,Z)dZ
\end{equation}

In order to calculate the total amount of acetylene produced by all the considered PAHs, it is necessary to take into account how PAHs are distributed by size. We adopt the dust size distribution ($dn/da$) from the work of \cite{wd01} (WD01) corresponding to $R_{\rm V}=3.1$. According to this distribution the number of all PAHs from 3 to 20~\AA{} is about $2 \cdot 10^{-7}$ with respect to H nuclei. If to adopt that a mean PAH contains about 50 atoms, the number of carbon atoms locked in PAHs is about $10^{-5}$ with respect to H nuclei. On an average, the abundance of carbon atoms with respect to H nuclei in gas is about $10^{-4}$,  therefore about 10\% of carbon atoms can be locked in PAHs, that may be significant to chemistry. We consider the size interval from $a_{\rm min}$ to $a_{\rm max}$, which correspond to the radii of the smallest (C$_{16}$H$_{10}$) and biggest (C$_{150}$H$_{30}$) PAHs in our sample. We divide this interval to $N_{\rm bin}$ bins with $N_{\rm bin}+1$ borders, $a^{\rm b}_j$. $N_{\rm bin}$ is taken to be equal 4. Each $j$th bin includes $N_{\rm PAH}^{j}$ PAHs, radius of which is in the size interval $(a^{\rm b}_j, a^{\rm b}_{j+1})$. The number of acetylene molecules in the $j$th bin is found as
\begin{equation}
N_{{\rm C}_2{\rm H}_2}^{j} = \sum\limits_{i=1}^{N_{\rm PAH}^{j}} N_{{\rm C}_2 {\rm H}_2}^{i}/N_{\rm PAH}^{j}
\end{equation}

The full number of acetylene molecules produced by all the considered PAHs with the adopted dust size distribution is estimated by
\begin{equation}
N_{{\rm C}_2{\rm H}_2} = \sum\limits_{j=0}^{N_{\rm bin}} \left(\frac{dn}{da}\right)_j N_{{\rm C}_2{\rm H}_2}^{j} (a^{\rm b}_{j+1} - a^{\rm b}_j)
\end{equation}
\label{total_number}

\begin{figure*}
	\includegraphics[width=0.45\textwidth]{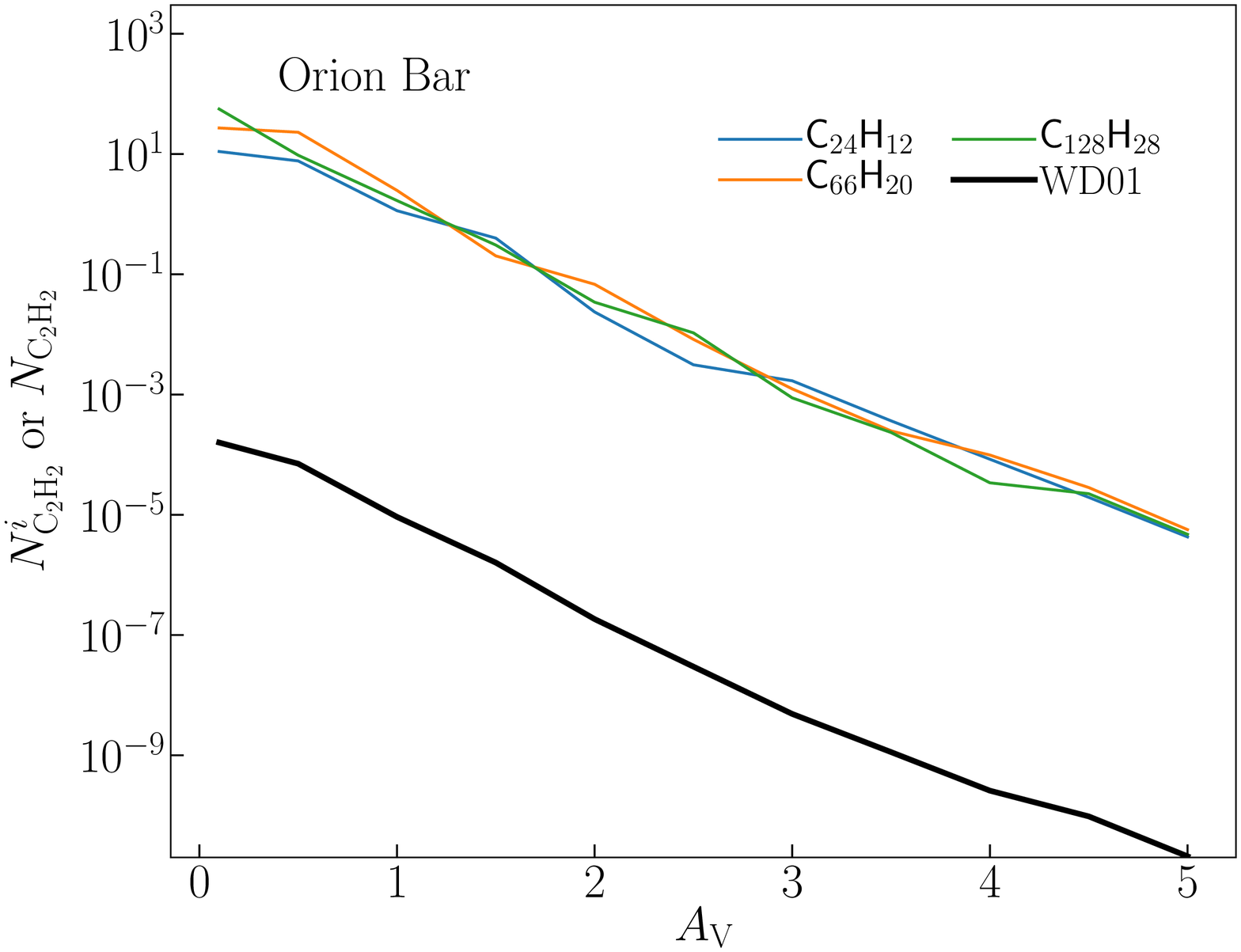}
	\includegraphics[width=0.45\textwidth]{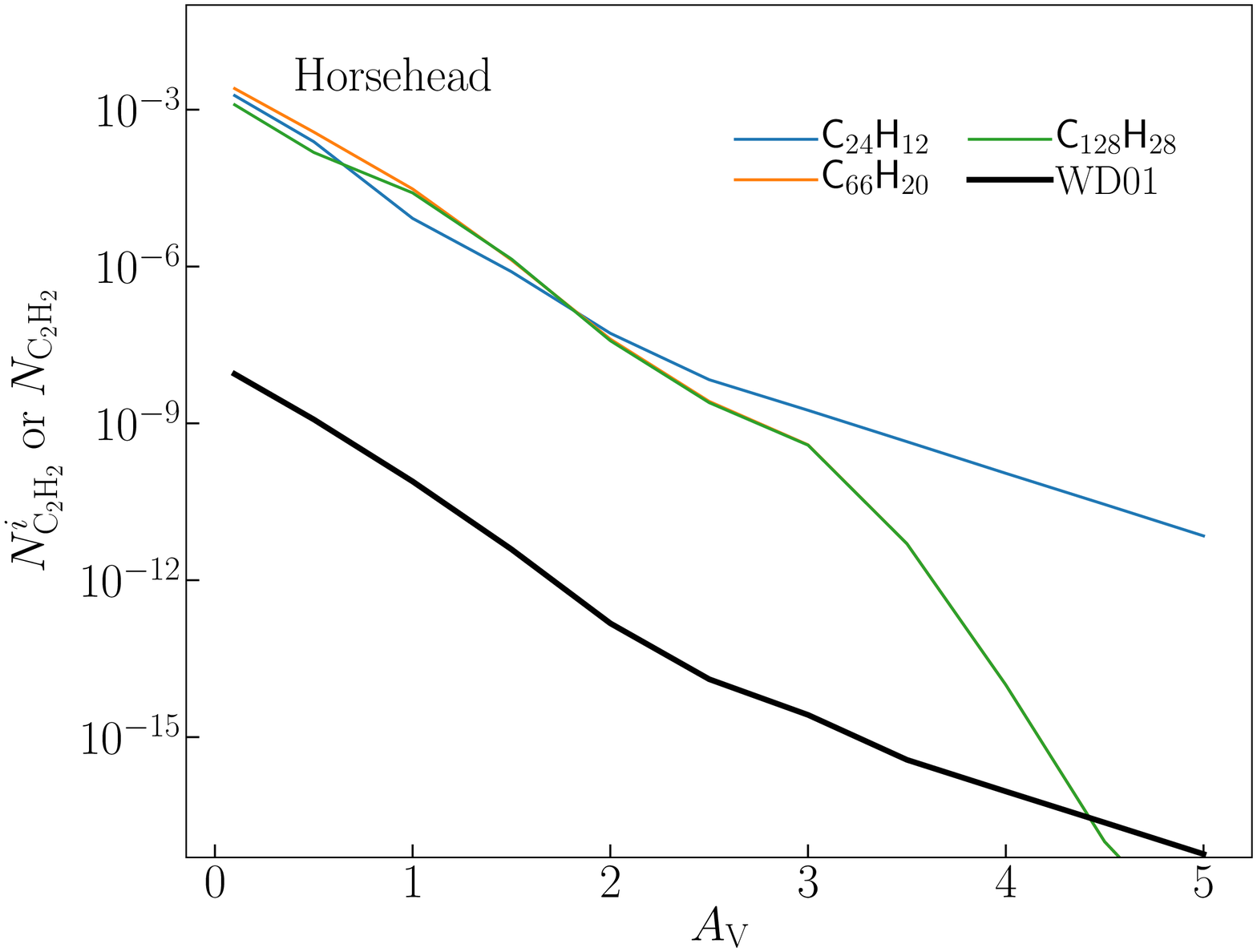}
	\caption{Dependence of the total number of the acetylene molecules produced by specific PAH molecules ($N_{{\rm C}_2{\rm H}_2}^{i}$) and all the considered PAHs with WD01 size distribution ($N_{{\rm C}_2{\rm H}_2}$) on the visual extinction in the Orion Bar (on the left) and the Horsehead nebula (on the right).}
	\label{nc2h2_av}
\end{figure*}

In Fig.~\ref{nc2h2_av} we illustrate the number of acetylene molecules produced throughout PDRs in the time period of 10$^6$~yr by three specific PAHs, C$_{24}$H$_{12}$, C$_{66}$H$_{20}$ and C$_{128}$H$_{28}$, and by all considered PAHs summed according to the WD01 size distribution. For both PDRs, the maximum amount of acetylene is produced at the minimum values of $A_{\rm V}$ (in our case, $A_{\rm V}=0.1$), where the radiation field intensity, the atomic hydrogen density and the temperature reach their maximum. The amount of acetylene produced by the destruction of PAHs decreases toward molecular clouds. In the Orion Bar the behaviour trends for all molecules are similar, while in the Horsehead nebula, the large PAHs have a sharp decline in the acetylene production at $A_{\rm V}\approx3$. Small PAHs are not as sensitive to the changes in external conditions, the amount of produced acetylene gradually decreases with an increase of $A_{\rm V}$, while large PAHs are hardly destroyed at some value of the radiation field intensity and produce acetylene molecules even less efficient than before.

It should be noted that the calculations for the rate of the PAH destruction may have uncertainties. Firstly, there are other PAH destruction models, different to the one used in this study. We calculate the rate of the dissociation describing the PAH molecules by using the microcanonical distribution. However, there is a number of other approaches~\citep{tielens05, montillaud13, allain96, lepage01} which can lead to different results, although, generally, the differences are not significant (see details of comparison in \cite{tielens05}). One of uncertainties of the results is in the approach chosen. Secondly, the process of the PAH destruction strongly depends on characteristics --- a PAH structure, size, presence of defective rings, etc. In the work of \cite{ekern98}, it was experimentally shown that there is no singly universal law by which all PAHs are destroyed, one can only reveal the main trends. Therefore, the results of the theoretical modelling of the PAH destruction may differ greatly from the experimental, true, values for the same molecule, since the model may not be applicable to this particular molecule. Thirdly, there is just little experimental data available for PAHs that are relevant for astrophysics, i.e. for PAHs with $N_{\rm C}>20-30$. The models are based on the results for small PAHs which have been extrapolated to larger sizes, and there is no way to verify the correctness of such extrapolation. Fourthly, we do not consider the process of `carbonization', that is the restoration of PAHs via reactions with carbon atoms or carbonaceous species. It was shown by \cite{allain96} that this process is ineffective against photodestruction, though our results can slightly change with taking into account it. Finally, we have considered only a set of compact PAHs, which are stable enough to survive in the ISM, however, the spectra of PDRs can not be well fitted only by this type of PAHs, other types, like catacondensed and irregular PAHs, are needed to explain the emission spectra~\citep{tielens13, andrews15}. If such PAHs exist in PDRs, they may contribute to small hydrocarbons as well. Thus, the results presented in this work are estimates which can be obtained from the modern models of the PAH destruction, but they have the uncertainties, which in some cases, especially for large PAHs, may be significant.   

\section{Gas phase chemical reactions versus PAH dissociation}
\label{chem_hydr_form}

\subsection{Acetylene formation in gas phase chemical reactions in PDRs}

To estimate the significance of PAH dissociation as a source of small hydrocarbons, it should be checked which rates of acetylene production are higher -- from the gas phase chemical reactions or from the PAH dissociation. Below we present the results of the calculations for the acetylene production rates in the gas phase chemical reactions for considered PDRs.

\begin{figure*}
	\includegraphics[width=0.45\textwidth]{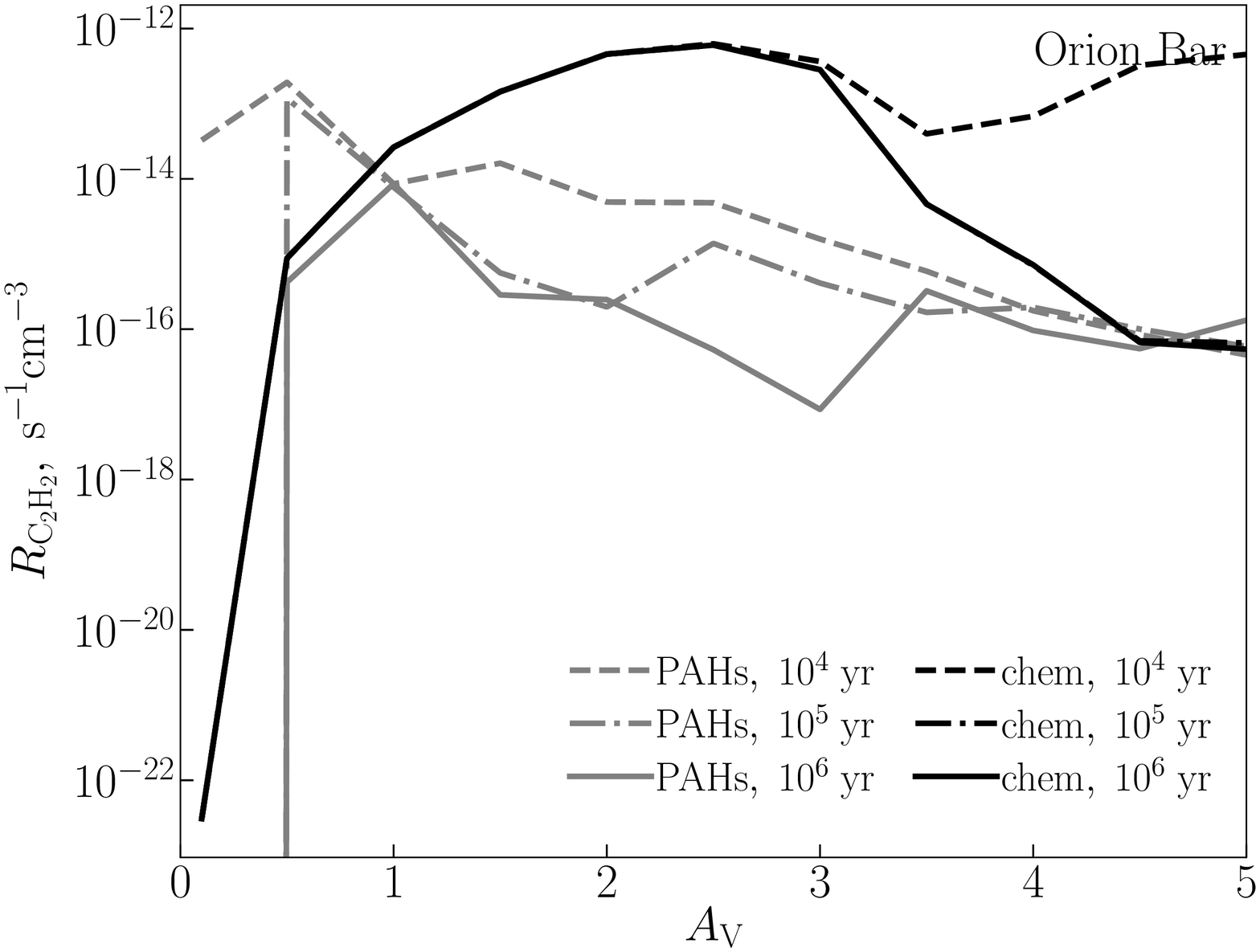}
	\includegraphics[width=0.45\textwidth]{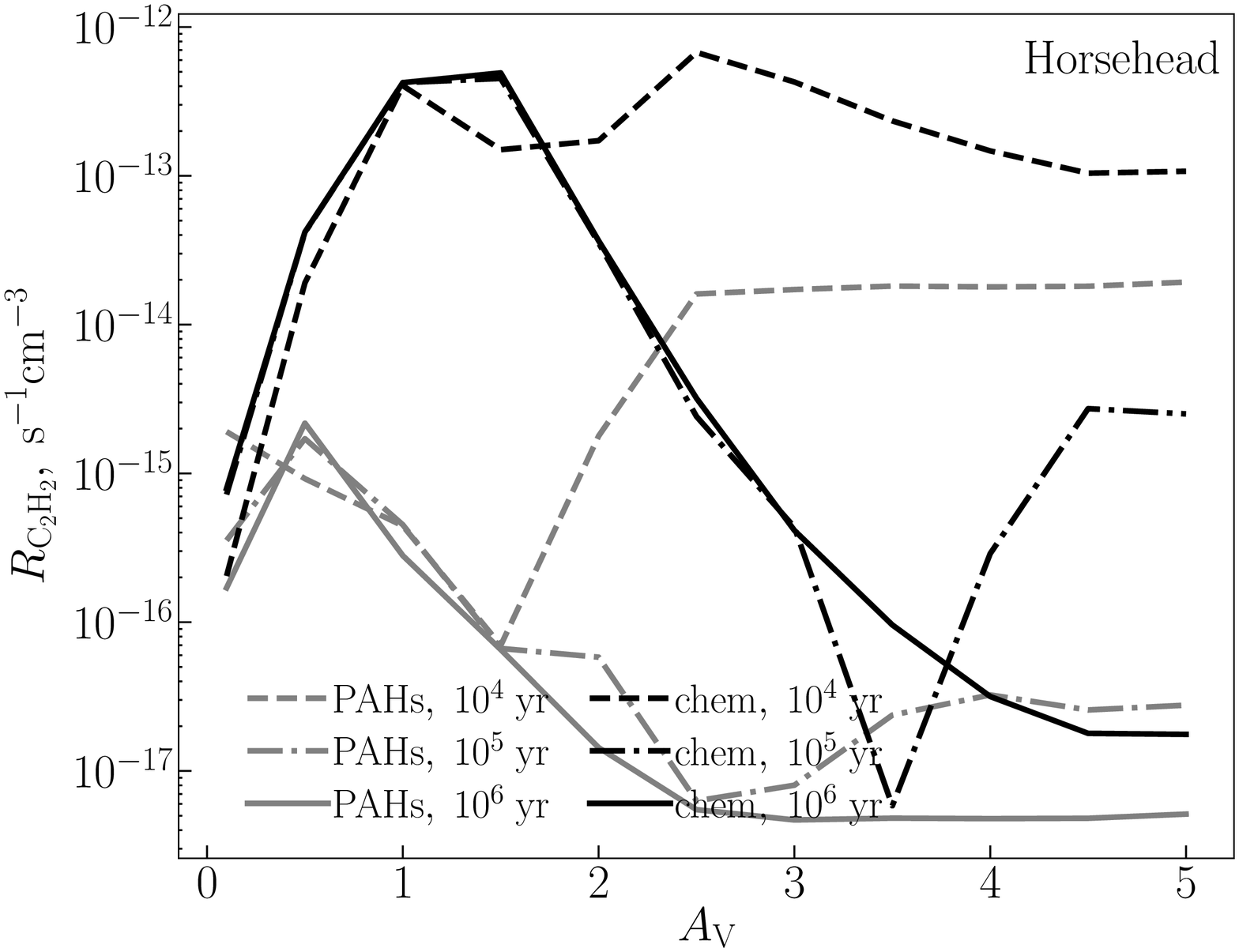}
	\caption{Rates of acetylene production via the PAH dissociation (gray color) and in gas phase chemical reactions (black color) in the Orion Bar (on the left) and in the Horsehead nebula (on the right). The calculations for the isochoric model of the Orion Bar are shown.}
	\label{compar}
\end{figure*}

We use the MONACO model~\citep{2013ApJ...769...34V, 2017ApJ...842...33V} to calculate the total rate of acetylene molecules formation in gas phase chemical reactions between atoms, molecules and ions, under the conditions illustrated in Fig.~\ref{profiles}. This model includes chemical processes occurring in the gas and on the surface of interstellar dust particles. Only diffusion chemistry is considered for the dust particles. The reactions occur between physisorbed atoms and molecules due to the Langmuir-Hinshelwood mechanism. Other  mechanisms (e.g. the Eley-Rideal one) are not considered. Before this work, the MONACO code was mainly applied to model chemical
evolution at the earliest stages of low- and high-mass star formation~\citep[e.g.][]{2013ApJ...769...34V, vasyunina14, ivlev15, rivilla16, 2017ApJ...842...33V, punanova18, nagy19}. Those regions are well shielded against stellar and diffuse interstellar UV radiation, and characterized by low gas and dust
temperatures. As such, although the network of chemical reactions utilized in the MONACO model 
is based on the OSU and KIDA networks and benchmarked with other codes for
the conditions described above, it does not include some processes
important for UV-dominated chemistry. In context of the scope of the
current study, perhaps, the most important class of processes currently
missing in our model are reactions with vibrationally excited H2~\citep{agundez10}. Those reactions were shown to be potentially important for chemistry of small hydrocarbons in highly FUV-irradiated
clouds such as the Orion Bar~\citep{cuadrado15}. Despite this limitation of our chemical model,
we believe that it is still suitable for the analysis of the role of PAH destruction in the formation of small hydrocarbons in PDRs. The current version of the model also does not include the formation of H$_2$ on PAHs. We present the results of the calculations from the MONACO model together with the rates of acetylene production via PAH destruction in Fig.~\ref{compar}. Note, that we find the total rate for PAHs analogously to the total number of produced acetylene molecules (Eq.~\ref{total_number}) taking into account the dust size distribution.

The rates in Fig.~\ref{compar} are given for three time moments at 10$^4$, 10$^5$ and 10$^6$~yr, as time goes on, they change. In the Orion Bar the rates of acetylene production by PAHs at 10$^4$~yr are slightly higher than the same rates at time moments 10$^5$ and 10$^6$~yr. This is due to decreasing of a number of available PAHs that can be destroyed under the certain conditions. At  $A_{\rm V}=0.1$ the PAH production stops after 10$^4$~yr due to full destruction of PAHs within this period. The PAH destruction is more efficient in production of acetylene than the gas phase reactions at 10$^4$~yr for  $A_{\rm V}<1$, at 10$^5$~yr -- in a range of $A_{\rm V}$ from 0.5 to 1, and  at 10$^6$~yr -- the PAH destruction is not more efficient than chemical reactions within the considered range of $A_{\rm V}$. The rates are comparable to each other at  10$^5$ and 10$^6$~yr at $A_{\rm V}>4$. If to consider another set of parameters (the isobaric model) and the corresponding calculations in Fig.~\ref{par_goi2}, then we can indicate that the PAH destruction is more efficient at $A_{\rm V}>3.5$ at any time. The last result arises from the high value of hydrogen density at these $A_{\rm V}$ relatively to the first model, which leads to possibility of recovery of the number of hydrogen atoms in PAHs (only PAHs with hydrogen atoms can produce acetylene).

In the case of the Horsehead nebula the rates for the PAHs are less than the rates for the chemical reactions for all the time moments and $A_{\rm V}$ excluding one point at $A_{\rm V}=0.1$, where the rates are comparable. In this  PDR the radiation field is weaker than in the Orion Bar and the PAH destruction is not so efficient. Only in the closest vicinity to the star, the PAH destruction may contribute the number of acetylene molecules comparable with the chemical reactions. 

\subsection{The contribution of PAH dissociation to the abundance of small hydrocarbons}

\begin{figure*}
	\includegraphics[width=0.45\textwidth]{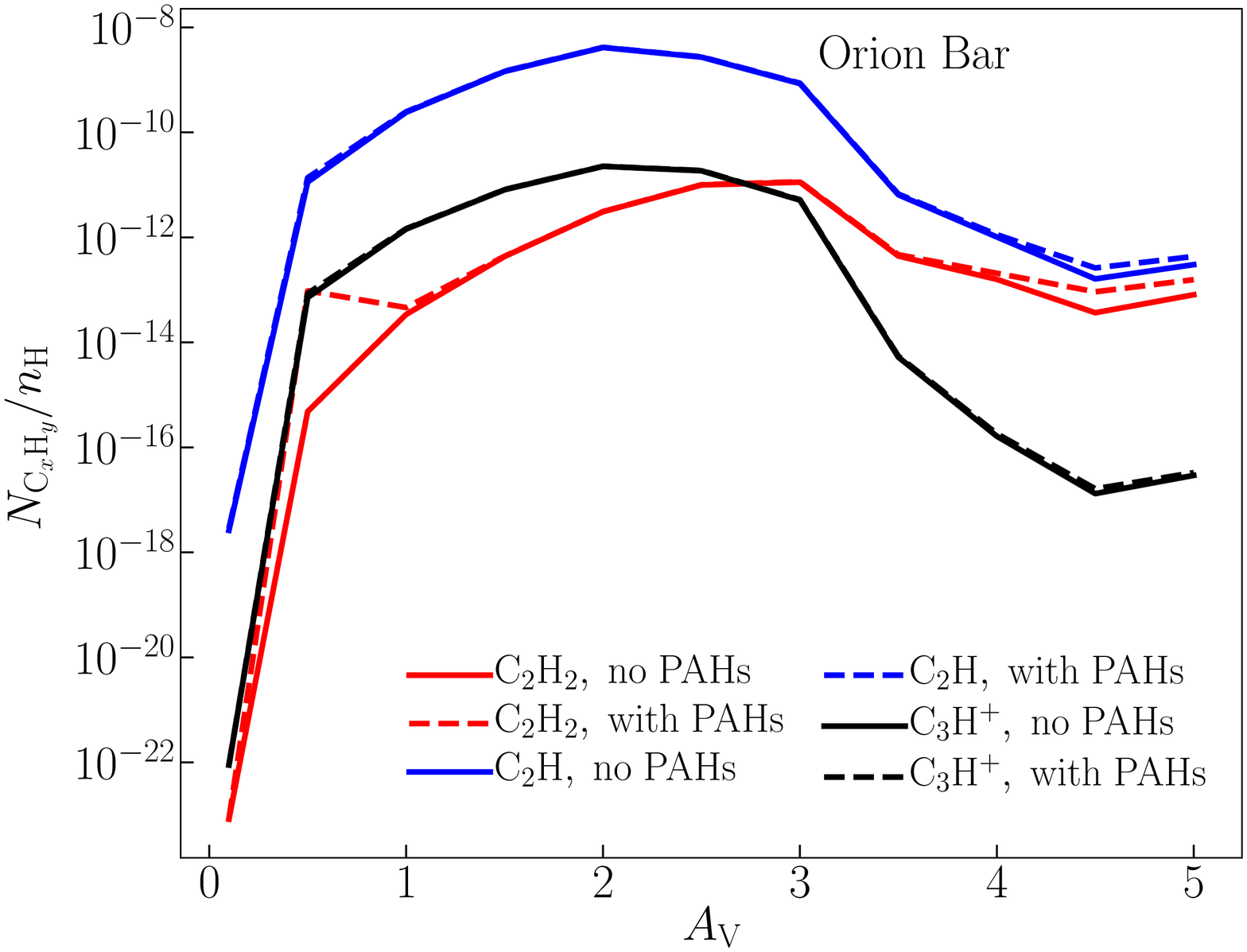}
	\includegraphics[width=0.45\textwidth]{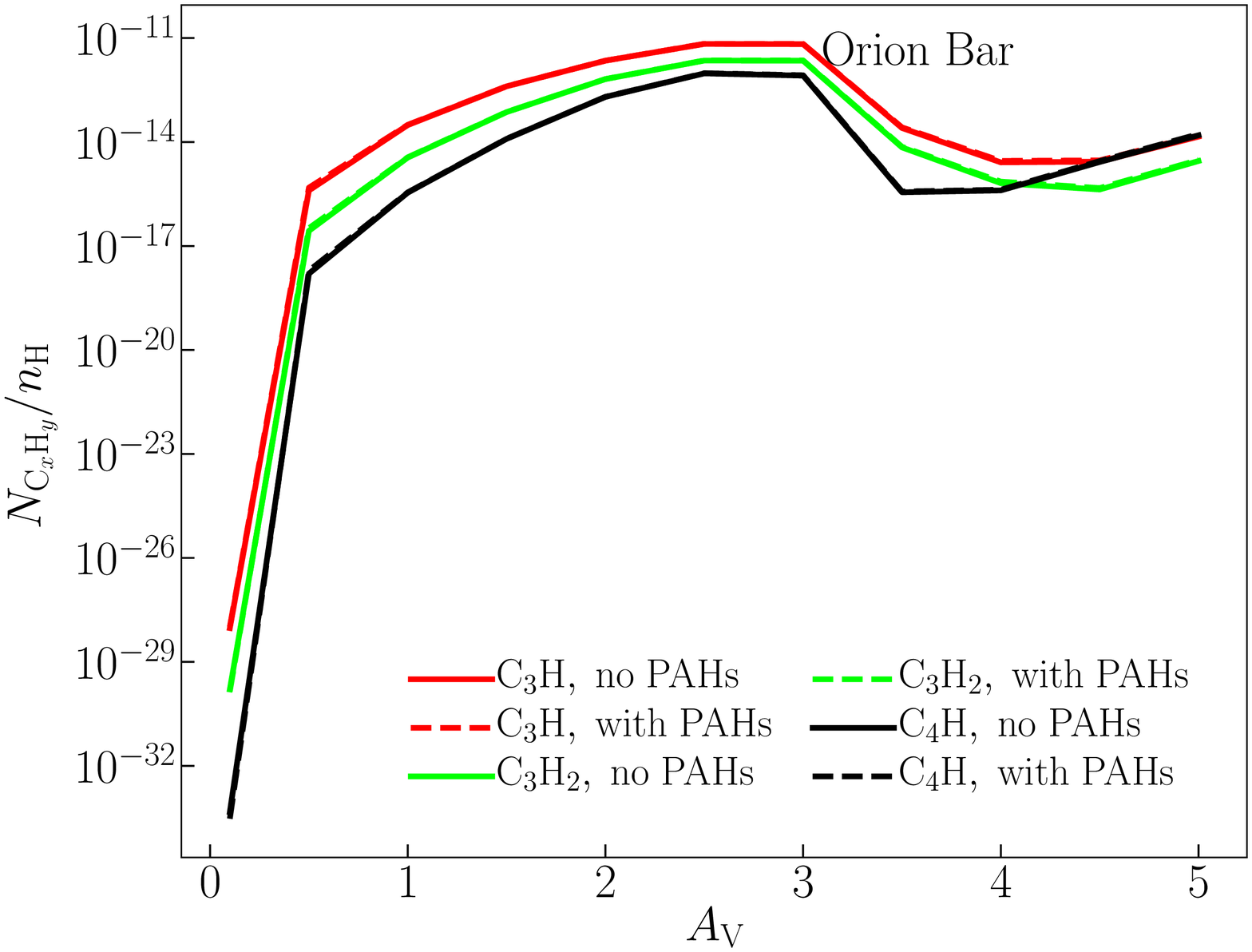}
	\caption{Abundance of small hydrocarbons produced in the gas phase chemical reactions without the contribution of PAHs (solid lines) and with PAHs (dashed lines) in the Orion Bar in approach of isochoric model.}
	\label{ncxhy_av_orion}
\end{figure*}

\begin{figure*}
	\includegraphics[width=0.45\textwidth]{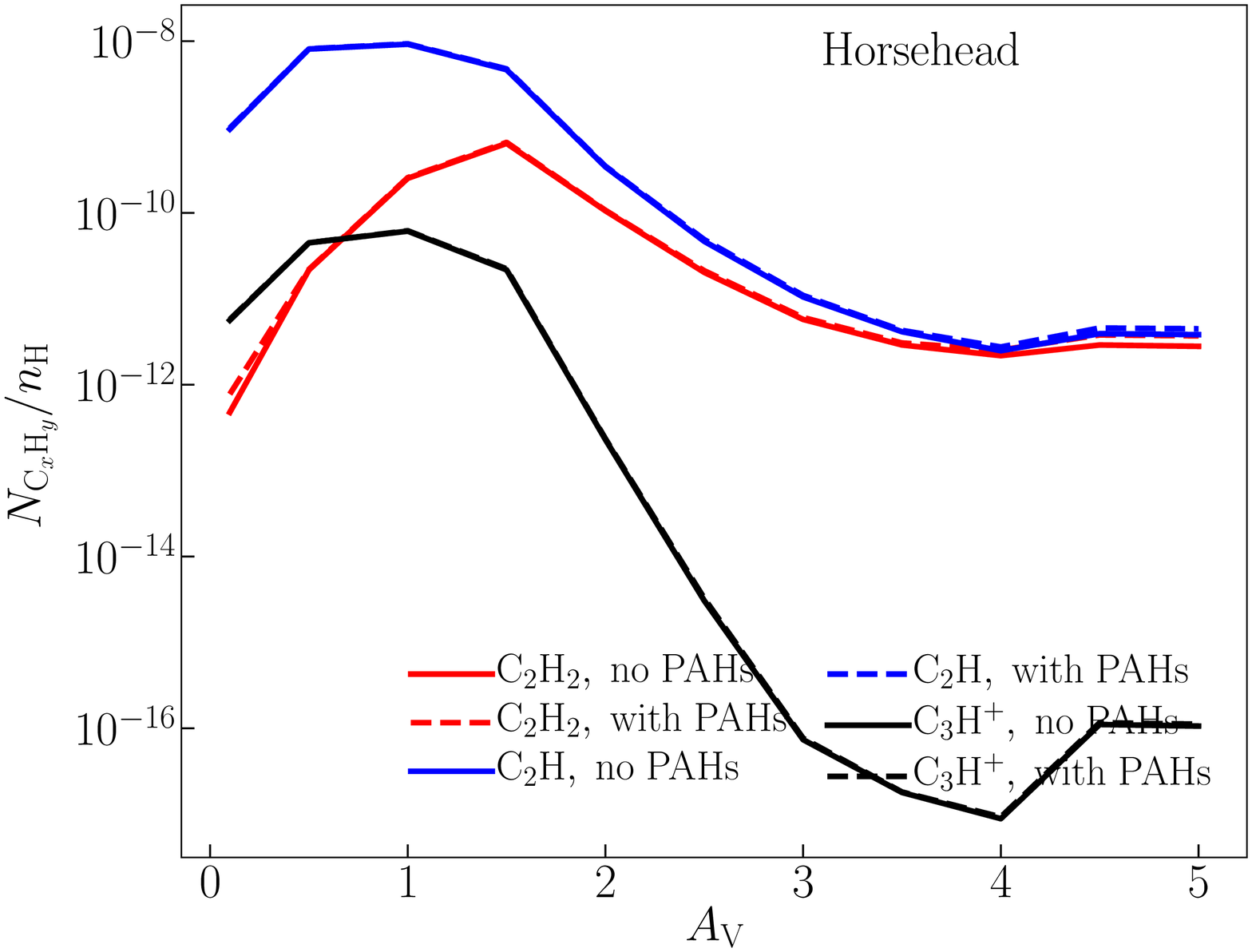}
	\includegraphics[width=0.45\textwidth]{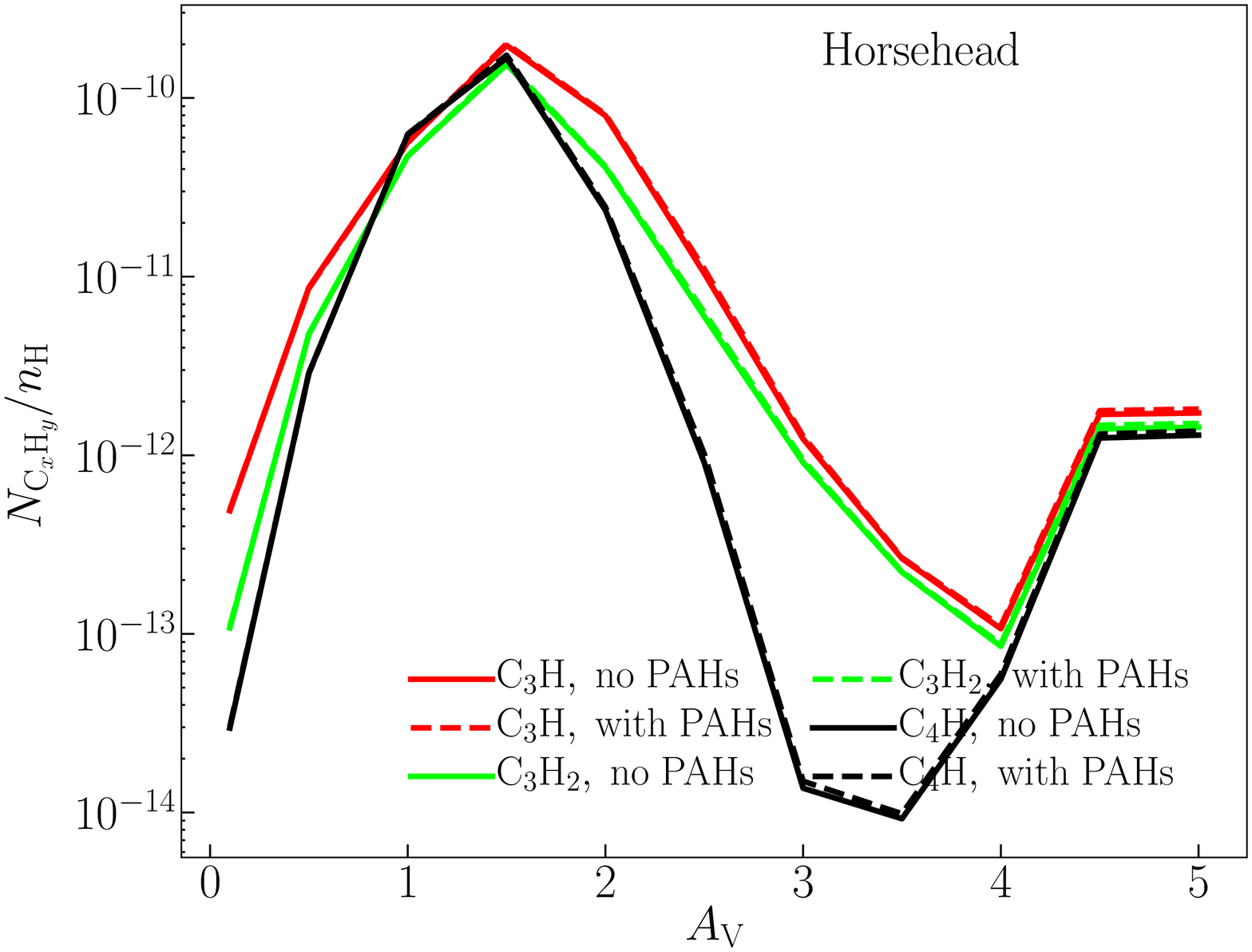}
	\caption{The same as in Fig.~\ref{ncxhy_av_orion} for the Horsehead nebula.}
	\label{ncxhy_av_horse}
\end{figure*}

Acetylene can participate in the reactions that form small hydrocarbons. Gas phase acetylene is destroyed in photodissociation reactions $\rm C_2H_2 + {\rm h}\nu \rightarrow C_2H + H$ and $\rm C_2H_2 + {\rm h} \nu \rightarrow C_2 + H_2 $~\citep{1984ApJ...282..172L}, as well as photoionization $\rm C_2H_2 + {\rm h} \nu \rightarrow C_2H_2^+ + e^-$~\citep{1971RvGSP...9..305H} under the powerful UV field conditions. The energy threshold of acetylene dissociation is 5.7~eV, the threshold of ionisation is 11.2~eV according to the data given by \cite{heays17}. The products of photodissociation and photoionization participate in other reactions that form different small hydrocarbons. Some of them are observed in both considered PDRs. Namely, they are  C$_2$H, C$_3$H, C$_3$H$^{+}$, C$_3$H$_2$, C$_4$H. Using the MONACO code without contribution of acetylene from the PAH dissociation and with this contribution we have estimated the abundance of these hydrocarbons in the Orion Bar and the Horsehead nebula. The results of these calculations are presented in Fig.~\ref{ncxhy_av_orion} and Fig.~\ref{ncxhy_av_horse} (and additionally, in \ref{chem_goi2}). The age of the Orion Bar is adopted equal to 10$^5$~yr~\citep{salgado16}, the age of the Horsehead nebula -- $5\cdot10^5$~yr~\citep{pound03}.

In the Orion Bar abundances of the considered small hydrocarbons are generally in consistency with the  abundances estimated in other works~\citep{cuadrado15, tiwari19}. The abundances  calculated with the PAH contribution and without PAHs are almost the same for all considered molecules except C$_2$H$_2$ at $A_{\rm V} \approx 0.5$ and $A_{\rm V} >4 $. It can be concluded that inflow of additional amount of C$_2$H$_2$ from PAHs by 2 orders higher than from chemical reactions at  $A_{\rm V}\approx  0.5$ does not change the abundance of other hydrocarbons substantially. However, a contribution of PAHs at $A_{\rm V} >3.5$ is seen in the calculations for the second model of the Orion Bar which are illustrated in Fig.~\ref{chem_goi2}. Thus, the PAH dissociation can play a larger role in the formation of small hydrocarbons relative to the gas phase chemical reactions at these $A_{\rm V}$ under the conditions predicted by the model in \cite{goicoechea19}. Nevertheless, basically we ascertain insignificance of PAHs as a source of small hydrocarbons in the Orion Bar.

Small hydrocarbons are observed in the Orion Bar. The peak emission of small hydrocarbons is shifted relative to the peak of the PAH emission~\citep{cuadrado15}: the peak of the PAH emission is at $A_{\rm V}\approx0.1$, while for small hydrocarbons the peak is at $A_{\rm V}>2$. These observations do not resolve spatially this emission peak, but it is most likely in the zone of molecular hydrogen. According to our results, the abundance of small hydrocarbons should have a peak around $A_{\rm V}\approx 1.5$. \cite{cuadrado15} have estimated the observational abundances of small hydrocarbons at $A_{\rm V}\approx1.5$. They vary from 10$^{-12}$ to 10$^{-8}$ for different species (the values are normalized to the hydrogen number density). Our abundances are slightly lower than the observational estimates, and the PAH contribution is negligible at this distance. Thus, despite the fact that the PAHs may contribute to the abundance of small hydrocarbons under some conditions, they can not provide the observable abundance at $A_{\rm V}\approx1.5$ where the observations are available. It is still necessary to have an alternative source. \cite{cuadrado15} suggested a scenario for efficient formation of small hydrocarbons -- endothermic reactions between C$^{+}$, radicals and molecular hydrogen which occur under the extreme conditions of the Orion Bar (high values of radiation field intensity). We do not consider these reactions, though we do not exclude that they have an important role in the production of small hydrocarbon. \cite{cuadrado15} emphasized that PAHs are not necessary to explain the observations, our results support this statement.

In Fig.~\ref{ncxhy_av_horse} the modelled abundances of small hydrocarbons are shown for the Horsehead nebula. As expected, taking into account the PAH contribution does not change the abundances. \cite{guzman15} have estimated the abundances of some small hydrocarbons in this PDR. At $A_{\rm V}\approx 1$ the abundance of C$_2$H is about $10^{-9}$, C$_3$H and C$_3$H$_2$ is about $10^{-10}$ and C$_3$H$^{+}$ is about $5\cdot10^{-11}$. The abundances obtained from the MONACO model are generally in consistency with these values. At $A_{\rm V}\approx 0.1$ all abundances become higher: C$_2$H -- $10^{-8}$,  C$_3$H$_2$ -- $10^{-9}$ and  C$_3$H$^{+}$ -- $10^{-11}$. The MONACO abundances do not achieve these values, though our results are consistent with the estimates from other works~\citep{pety05, guzman15}. Consideration of PAH destruction does not improve the modelling. In this PDR, the peak of emission of small hydrocarbons (namely C$_3$H$_2$, C$_2$H, C$_4$H) coincides with the peak of the PAH emission based on observations ($A_{\rm V}\approx 0.1$),  that is why it was assumed that the PAH destruction provides these hydrocarbons~\citep{pety05, guzman15}. According to our calculations, the contribution of the PAH destruction to the abundance of small hydrocarbons is negligible.  The radiation field intensity is not enough to destroy large PAHs, it is sufficient only for destruction of the smallest PAHs. The dehydrogenation of such PAHs occurs quickly. The recovery of the number of hydrogen atoms does not occur efficiently. As a result, only a few acetylene molecules are produced within the time period when the PAHs have hydrogen atoms. Therefore, PAHs contribute a low amount of additional acetylene to chemical reactions. Thus, we disprove the assumption about the efficient PAH contribution to the small hydrocarbons abundance.

Apart from the detailed calculations of the PAH dissociation, let us make some analytical estimates. As we observe PAHs in PDRs, than at least some of them should survive about 10$^5-$10$^6$~yr. If assumed so: 1) the number of PAHs decreases by the exponential law $n_{\rm PAH}=n_{\rm PAH}^{0}\exp(-kt)$, where $k$ is the rate of their destruction; 2) the number of PAHs decreases by $e$ times for 10$^6$~yr; then $k$ is about 10$^{-14}$-10$^{-13}$~s$^{-1}$. These rates are too low relative to the typical rates of chemical reactions which play a role in chemical evolution (10$^{-10}$-10$^{-9}$~s$^{-1}$). From this point of view, PAHs can not contribute a lot to the chemical evolution. Of course, the above rates are average. Small PAHs are destroyed much faster (at least in the Orion Bar), while large PAHs survive and may live longer than the PDRs themselves. But if small PAHs dissociate very fast, then they contribute only in the beginning, after that they can not be a prominent source of acetylene, because only large and stable PAHs remain which dissociate slowly. Thus, from this point of view, despite the high rates of destruction of small PAHs, they play a role in the chemical evolution only for a short period. To be able to make them important for a lifetime of a PDR, there should be a source of PAHs, which supplies the PAH abundance. The possible source of PAHs can be evaporating very small grains~\citep{pilleri12}. 

It should be noted that our calculations of the PAH contribution depend on the adopted dust size distribution. We take the one from the work of \cite{wd01} which corresponds to the parameter $R_{\rm V}=3.1$ with the maximum number of PAH-size grains. This distribution successfully describes the observed extinction curve and infrared emission in different parts of the Milky Way. But we assume that the size distribution in the considered PDRs (especially in molecular cloud) may differ from the adopted one. For example, $R_{\rm V}$ is estimated as $4-5$ in the Orion Bar~\citep{bohlin81}. It means that the abundance of PAHs is overestimated in our calculations, but even in this case we can not explain the observations. In fact, we need to have the abundance of PAHs by several orders higher, which is impossible to get varying the size distribution.

We also remark that due to the scarcity of the experimental data, in our model we consider only acetylene among carbonaceous products of the PAH dissociation. But this is plausible only for the PAHs in the normal and dehydrogenated states, while in the super-hydrogenated states other products are also possible (including the observable ones). Perhaps, the inclusion of other products in the model will increase the abundance of small hydrocarbons as well, however, we do not expect that this growth will be enough to explain the observations. 

Aside from PAHs, small hydrocarbons can be produced by large dust grains, if the material of these grains is hydrogenated amorphous carbon~(HAC). To date, many experiments have been carried out, which indicate that the formation of carbonaceous dust with a disordered structure (such as that of HAC) is most likely under ISM conditions~\citep{jager09}, and that the HAC grains fragment into different small hydrocarbons (including PAHs and acetylene) under the influence of the UV radiation field~\citep{scott97, alata15, duley15}. \cite{alata15} applied the laboratory results of HAC photodestruction coupled with a PDR model to the Horsehead nebula, and they concluded that HACs could provide the amount of small hydrocarbons needed to explain the observations of this PDR.

\section{Conclusions}\label{conclusions}

\begin{itemize}
\item{
We applied the model of the evolution of PAHs to the conditions of the high radiation field intensity. Two PDRs were studied: the Orion Bar and the Horsehead nebula. For these PDRs we took sets of parameters modelled in other works. For the Orion Bar two models, isochoric and isobaric, were considered. It was shown that the PAH size and charge, the radiation field intensity, the number density of the hydrogen atoms, ions, and electrons, and gas temperature affect the rate of PAH photodestruction. The numbers of the acetylene molecules produced by a specific PAH molecule for a lifetime of PDRs can differ by several orders for different conditions and charges. We calculated the evolution of PAHs throughout PDRs from the star to the molecular cloud at up to $A_{\rm V} = 5$. It was found that in the considered range of distances, PAHs produce the acetylene molecules the most efficiently in the vicinity of ionized stars, at $A_{\rm V} \approx 0.1$.}

\item{
The rates of the acetylene production via the PAH destruction and in the gas phase chemical reactions were calculated and compared to each other. It was shown that the rate of the acetylene production in gas phase chemical reactions is lower than the same rate for the PAH dissociation in the Orion Bar by up to $A_{\rm V}<1$ during 10$^4$~yr and around $A_{\rm V}\approx 0.5-1$ during 10$^5$~yr for both considered sets of parameters. PAHs may produce acetylene molecules more efficiently at $A_{\rm V}>3.5$ in approach of the isobaric model. In the Horsehead nebula, the gas phase chemical reactions dominate in the production of acetylene at a whole considered range of $A_{\rm V}$.}

\item{
Using the chemical model MONACO we estimated the abundances of the observed small hydrocarbons (C$_2$H,  C$_3$H, C$_3$H$^{+}$, C$_3$H$_2$, C$_4$H) in PDRs with and without inclusion of the PAH contribution to the acetylene abundance. The peak of abundance of the small hydrocarbons is around $A_{\rm V}\approx1.5$. In approach of the isochoric model, the PAH dissociation increases the modelled abundance of acetylene in the Orion Bar at $A_{\rm V}\approx0.5$ and  $A_{\rm V}>4$, but it almost does not change abundances of other small hydrocarbons. In approach of the isobaric model, increasing of all considered hydrocarbons occurs at $A_{\rm V}>3.5$ due to the PAH dissociation. PAHs does not affect the results in the Horsehead nebula. The calculated abundances are lower than the observational values at $A_{\rm V}\approx1.5$ in the Orion Bar and at $A_{\rm V}\approx1$ in the Horshead nebula even when the PAH destruction is taken into account. Thus, at least in our model we conclude that the PAHs are not a major source of small hydrocarbons, which were recently observed to have high abundances towards the Orion Bar and the Horsehead nebula by \cite{pety05, cuadrado15}.}
\end{itemize}

The work was supported by the grant of the Russian Science Foundation (project 18-12-00351). We thank Javier Goicoechea who provided the profiles of parameters in the Orion Bar. We are grateful to the anonymous referee for constructive suggestions that allowed us to improve the manuscript.

\section*{Data availability}

The data underlying this article are available in Figshare at {\tt\url{https://doi.org/10.6084/m9.figshare.12597092.v1}}.

\bibliographystyle{mnras} 
\bibliography{refs_acetylene}

\appendix
\section{The Orion Bar: the results for the isobaric model}
\label{app}
\begin{figure*}
	\includegraphics[width=0.45\textwidth]{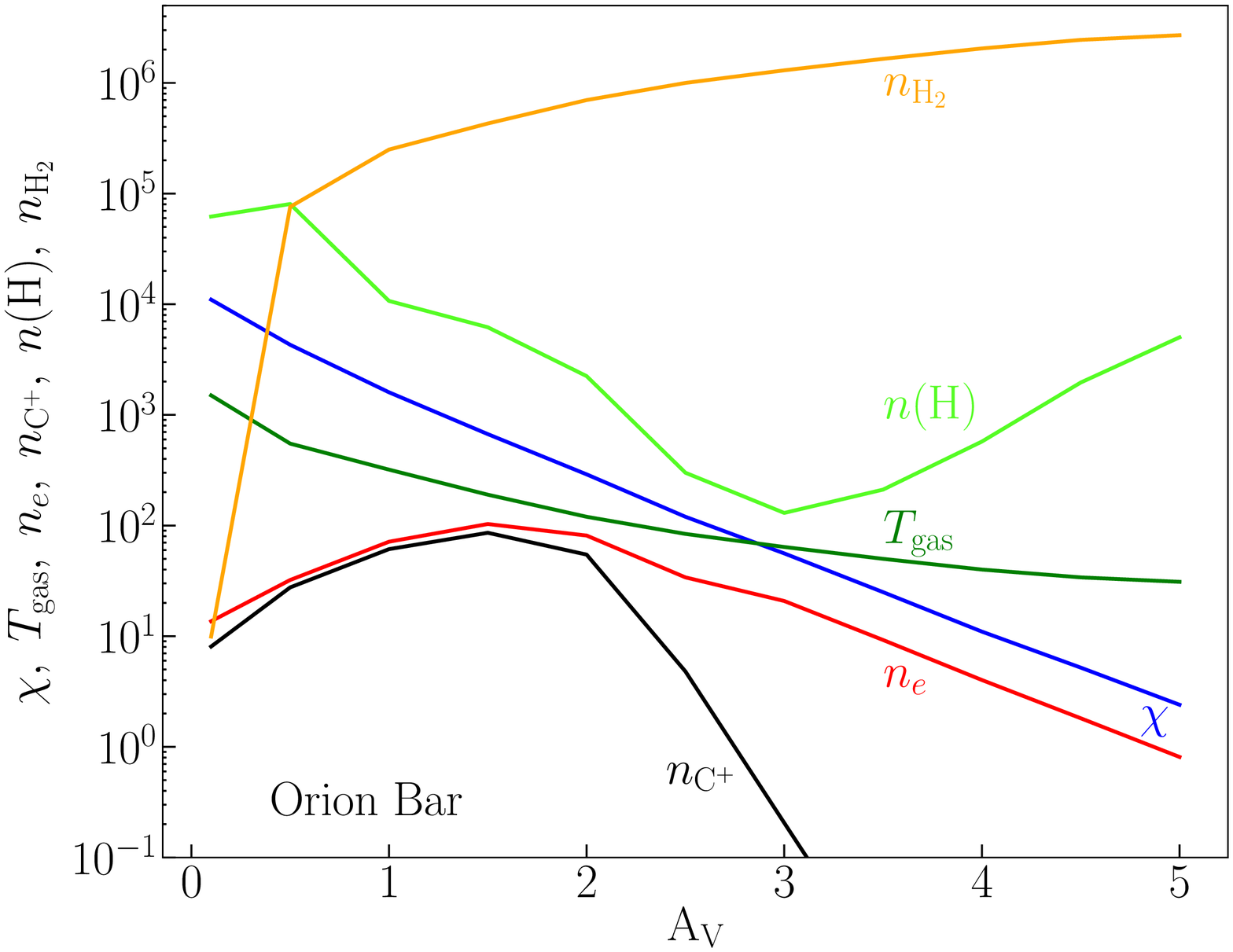}
	\includegraphics[width=0.45\textwidth]{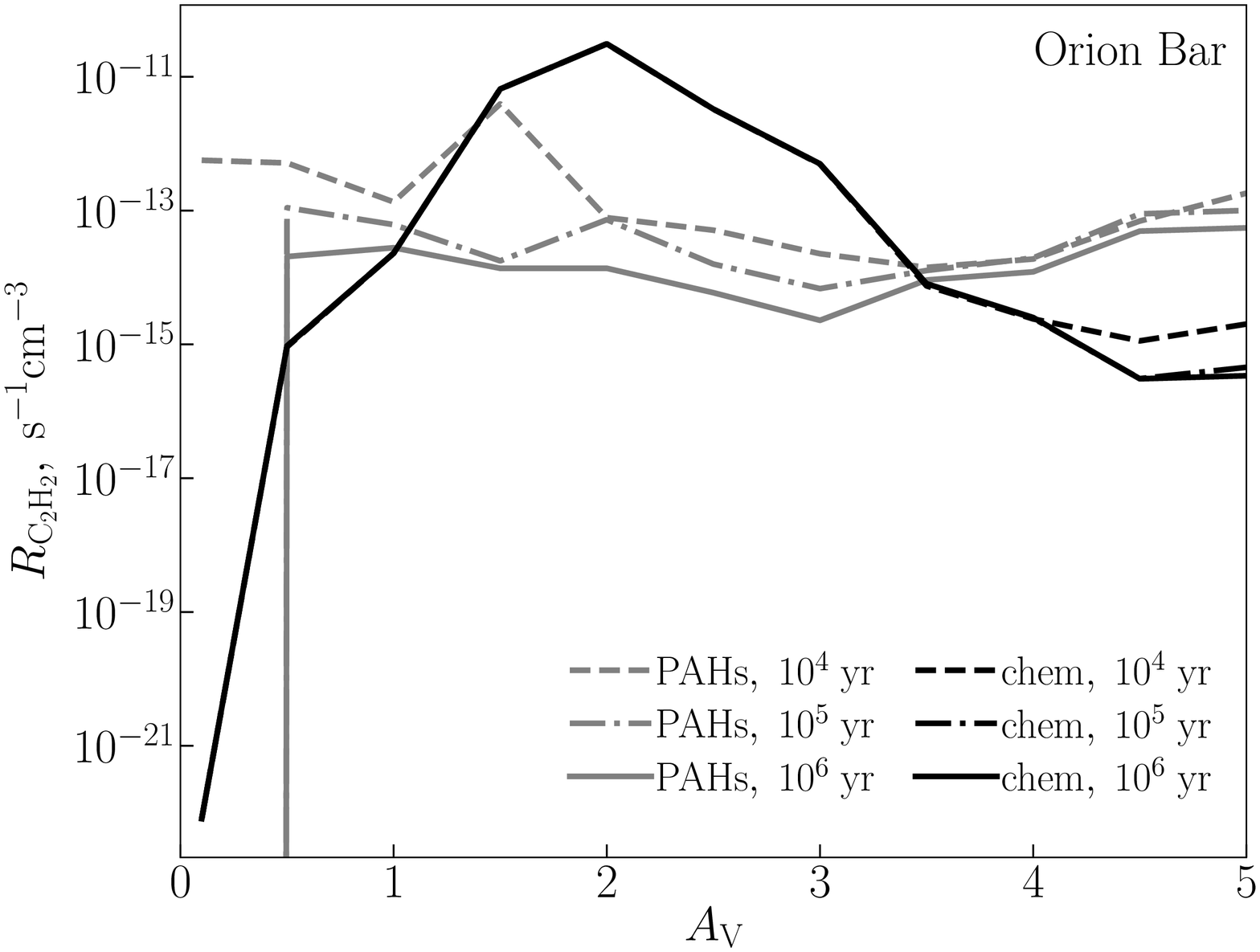}
	\caption{In the left: parameters of the Orion Bar from the isobaric model~\protect\citep{goicoechea19}. In the right: the same in Fig.~\ref{compar} for the isobaric model of the Orion Bar.}
	\label{par_goi2}
\end{figure*}

\begin{figure*}
	\includegraphics[width=0.45\textwidth]{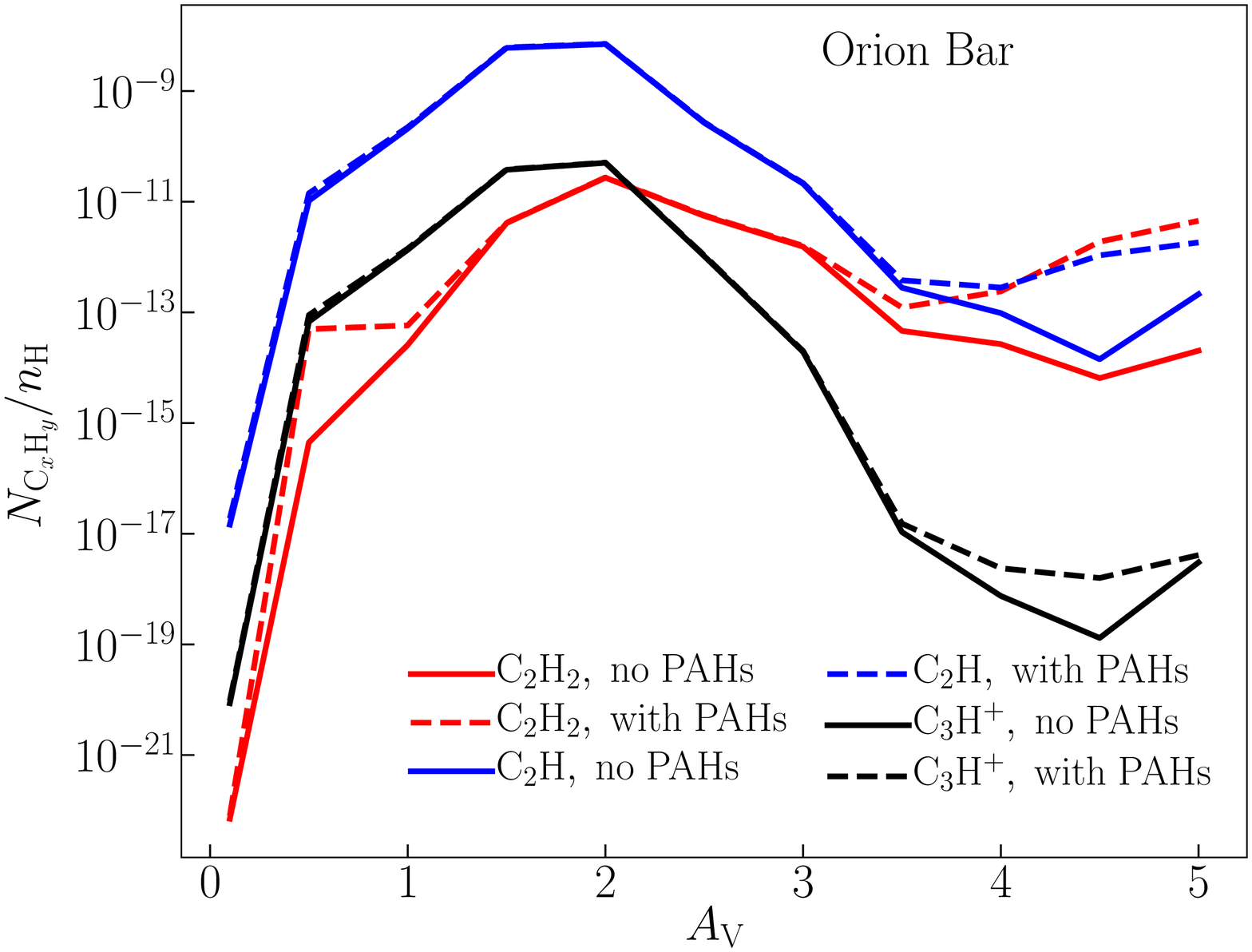}
	\includegraphics[width=0.45\textwidth]{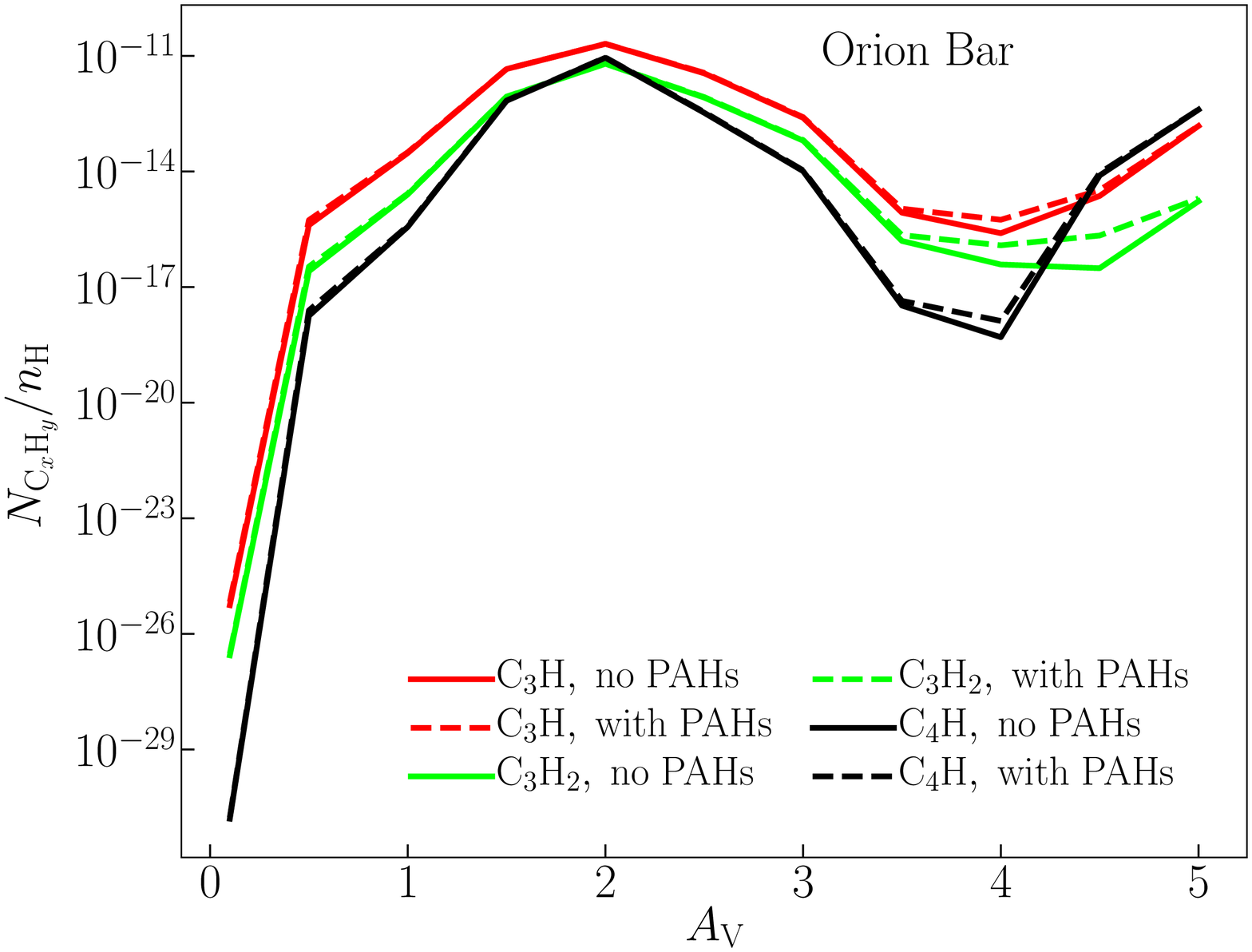}
	\caption{The same as in Fig.~\ref{ncxhy_av_orion} for the isobaric model of the Orion Bar.}
	\label{chem_goi2}
\end{figure*}

\bsp	
\label{lastpage}
\end{document}